\documentclass[a4paper,12pt]{article}
\usepackage{amsmath,amssymb}
\usepackage[english]{babel}
\usepackage{graphicx}
\usepackage{color}

\title{Electroconvective instability of self--similar equilibria}

\author{E. A. Demekhin $^{[1]}$, S. V. Polyanskikh\\
Kuban State University\\
Stavropolskaya st., 149\\
Krasnodar 350040, Russia,\\
and Yury Shtemler\\
Dept. of Mechanical Engineering
Ben Gurion University of the Negev\\
P.O. Box 653, Beer Sheva 84 105, Israel}

\begin{document}

\setlength{\baselineskip}{18pt} \maketitle
\begin{center}
Abstract
\end{center}
{\color{black} Stability of electro--hydrodynamic processes between
ion--exchange membranes is investigated. Solutions of the
equilibrium problem are commonly described in the one--dimensional (1D) steady--state approximation.
In the present work, a novel class of 1D unsteady self--similar
equilibrium solutions is developed, whose existence is supported
by  recent experiments. 1D unsteady two--parametric family
of self--similar equilibrium solutions and their stability are
studied both asymptotically in small dimensionless Debye length,
and numerically. The self--similar solutions and marginal stability
curves obtained in both approaches are in a fair agreement with each
other at intermediately large times for which  the dimensionless
distance between membranes is large.

\footnotetext[1] {E-mail address: edemekhi@gmail.com}

\newpage
\setlength{\baselineskip}{24pt}
\renewcommand{\thesection}{\Roman{section}{.}}
\section{INTRODUCTION}
\renewcommand{\thesubsection}{\Alph{subsection}{.}}

Problems of electro--kinetics attract a great attention due to
 a rapid development of micro--, nano-- and biotechnologies.
Among  numerous modern applications of  electro--kinetics,  note
micro--pumps and biological cells, electro--polishing of mono-- and
poly--crystalline aluminum, growth of aluminum oxide layers  for
creating micro-- and nano--scale regular structures, such as quantum
dots and wires.

Study of the space charge in the electric double--ion layer (EDL)
is a fundamental problem of modern physics
\cite{Smr1}--\cite{Chu1}, first addressed by Helmholtz.  A detailed
description of the current state of the art is presented by Zaltzman
and Rubinstein \cite{Rub13}. As distinct from the commonly
accepted stability studies of the steady--state equilibria, the
present work describes a novel class of unsteady transient
equilibria predicted numerically (Demekhin et al. \cite{Dem}) and
observed experimentally (Yossifon and Chang \cite{YCh}). Since the
present approach is strongly based on the principal results
developed for steady--state equilibria, they are discussed below
along with  basic experimental data.

A typical experimental voltage--current (V--C) curve in Figure~\ref{P2} gives a first
insight into the system. The V--C curve depicts three characteristic
regimes: the regime A of  under--limiting currents proportional
to a low potential drop; the regime B of limiting currents (the plateau region), and the regime C of
over--limiting currents proportional to the potential drop as in the regime A.

\begin{figure}[htb]
  \begin{center}
    \includegraphics[height=6cm]{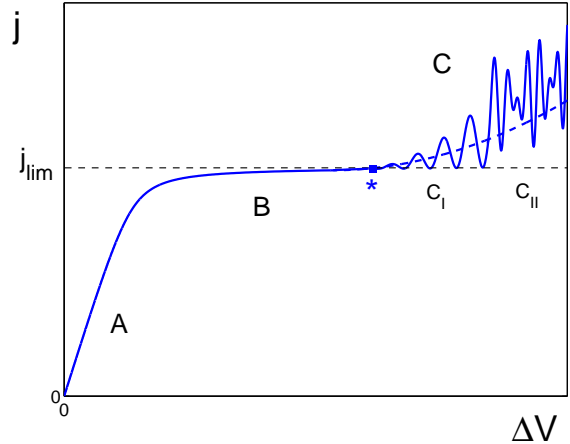}
  \end{center}
\caption[]{Schematic V--C curve for ion--exchange  membranes. A, B and C
represent  under--limiting,  limiting and over--limiting regimes,
respectively;
* depicts the threshold of the
over--limiting regimes; $C_I$ and $C_{II}$ are the regions of
regular and irregular current oscillations; $j$ and $j_{lim}$ are
the current density at the bottom membrane and
its limiting value; $\Delta V$ is the potential drop.
}\label{P2}
\end{figure}

\textit{Theoretical description of 1D steady--state equilibria.}
Early studies assume (see, e.g., \cite{Lev}) that outside of a small
vicinity of the membrane surface $y=0$, right away after EDL,
there is an equilibrium diffusion layer which satisfies the
condition of electro--neutrality. For 1D steady--state  equilibria
homogeneous along  membranes, this yields the following solution:
\begin{equation}\label{eq3015}
\displaystyle c^+=c^-=\frac{j}{j_{lim}}(y-1)+1, \ \ \ \Phi=\ln\frac{j(y-1)+j_{lim}}{j_{lim}-j},\ \ \
\displaystyle\Delta V=\ln\frac{j_{lim}}{j_{lim}-j}, \ \ \ j_{lim}=2.
\end{equation}
Here $c^{\pm}$ are ion concentrations,  $0\leqslant y\leqslant 1$ is the coordinate
normal to the membrane surfaces, $j$,  $j_{lim}$ and $\Delta V$
have the same meaning as in Fig.~\ref{P2}. The solution
\eqref{eq3015} is made dimensionless using the diffusion layer
thickness, the bulk ion concentration and the thermodynamic
potential as the characteristic scales.
 As follows from \eqref{eq3015},  the V--C curve  obeys a
 linear Ohmic relationship for small $j$ (region A in Fig.~\ref{P2}).
 As it was first pointed out by Levich \cite{Lev},  the solution \eqref{eq3015}
 has no physical meaning for
  $j\geqslant j_{lim}$.

Several works (Smyrl and Newman \cite{Smr1}, Rubinstein and
Shtilman \cite{Rub1}, Grafov and Chernenko \cite{Gr1}) are focused
to remedy this inconsistence. Rubinstein and Shtilman \cite{Rub1}
came up with  ideas of the non--equilibrium nature
of EDL and of the extended space charge (ESC) region
which is much thicker than EDL. Besides this, the fundamental novelty of this study is
the emergence  of the diffusion layer near a
permselective solid/liquid interface for the limiting and
overlimiting currents. This becomes possible due to a lucky choice
of a suitable  model problem.  The quiescent steady--state 1D
problem has a small parameter $\varepsilon$ (dimensionless Debye
length) at the highest derivative of the Poisson equation for the
electric potential. The nonlinear problem is reduced to the
Painleve equation of the second kind for the electric field and
solved numerically. The work \cite{Rub1} gives a key to the
understanding of the limiting regimes and provides the success of
their further studies \cite{UrtKir}--\cite{Urt3}.

 An effective asymptotic approach in small $\varepsilon$ (the decomposition method)
 yields the following simple analytic solution of the 1D quiescent steady--state
 problem \cite{UrtKir}--\cite{Urt3}:
\begin{equation}\label{eq3020}
c^{+}= c^{-}= 0,\ \ \ F = \Delta F \left[1-\left(1-\frac{y}{y_m}\right)^{3/2}\right],\ \ \ 0\leqslant y\ < y_m,
\end{equation}
\begin{equation}\label{eq3021}
c^{+}= c^{-}= \frac{j}{j_{lim}}(y-y_m),\ \ \ F  = 0, \ \ \ y_m\leqslant y\ < 1,
\end{equation}
where the V--C characteristics  and the thickness of the ESC
region~$y_m$ are
\begin{equation}\label{eq3022}
\Delta F = \frac{4}{3}\sqrt{\frac{j}{j_{lim}}}\ y_m^{3/2},\ \ \
y_m = 1-\frac{j_{lim}}{j},\ \ \ j_{lim}=2.
\end{equation}
Here $F \equiv \varepsilon V$, $\Delta F \equiv \varepsilon
\Delta\Phi$. The solution \eqref{eq3020}--\eqref{eq3022} may be
additionally simplified for the near--critical values of $y_m$ and
$\Delta F$. These values are small at small current deviations
$(j-j_{lim})$, which in turn may be ordered in $\varepsilon$.
The relations \eqref{eq3020} and
\eqref{eq3021} determine, to leading order in $\varepsilon$, the
outer solution of the problem in the space--charge and
electro--neutral regions, respectively. In general,  matching
conditions should relate the outer solution with the inner
solutions in the thin internal boundary layer at $y{\,=\,}y_m$
and in the double--ion boundary layer near $y{\,=\,}0$. Here the
terms ``outer" and ``inner" solutions have the usual meaning of
the method of
 matched asymptotic expansions  \cite{VanDyke1964}. Instead,   patching
conditions are applied to the outer solution at $y=y_m$ and $y=0$.
This yields a fair approximation of the problem solution in the
outer regions   applying the patching conditions to  the outer
solutions and ignoring inputs of the inner solutions in the
boundary layers. This is qualitatively illustrated in
Fig.~\ref{P3}, where the electric field $E$, the charge density
$\rho{\,=\,c^+\,-\,}c^-$ and the concentrations $c^\pm$ versus
coordinate $y$ are  depicted for both the asymptotic and exact
solutions at  small $\varepsilon$.

\begin{figure}
  \begin{center}
   \includegraphics[height=4.2cm]{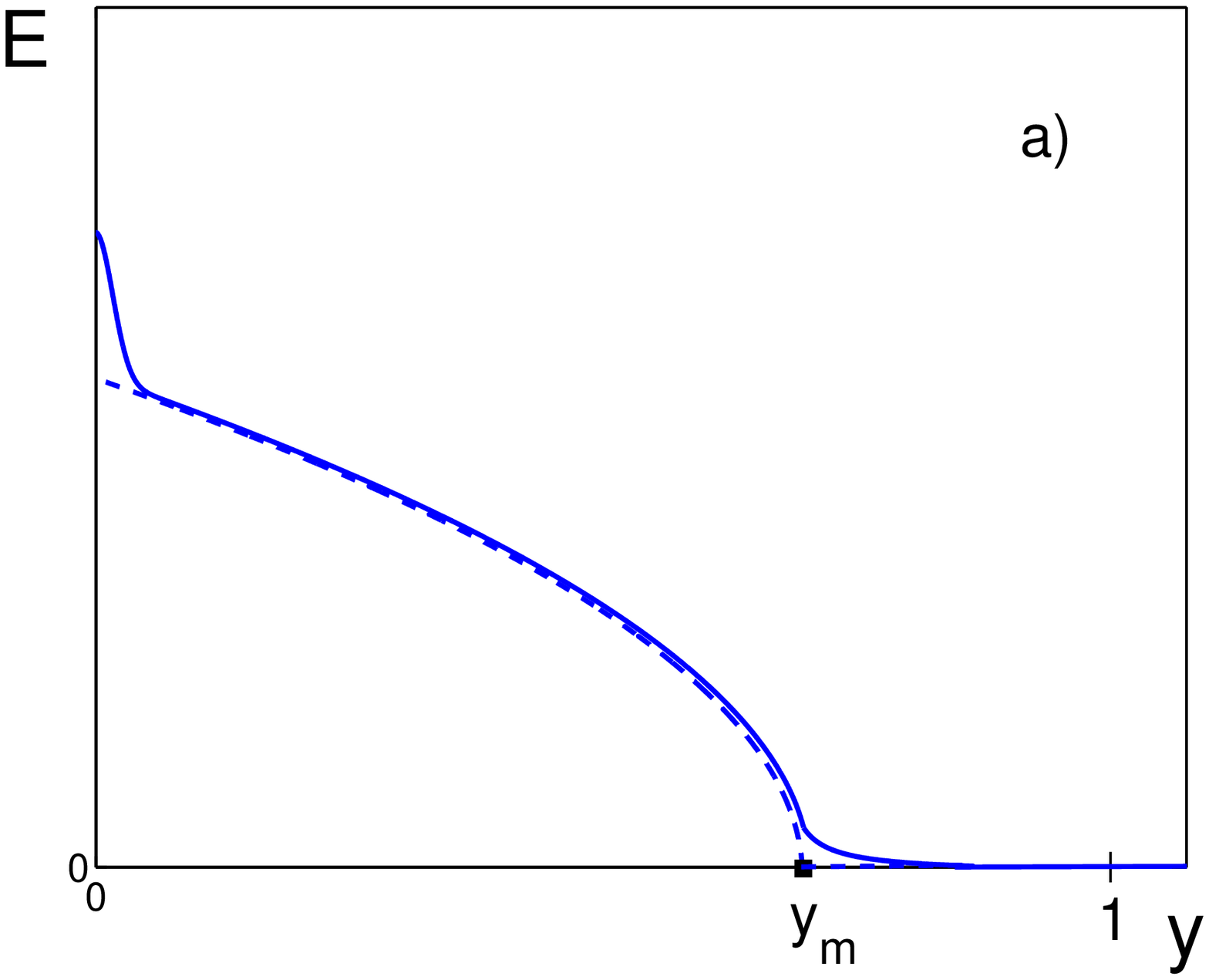}
   \includegraphics[height=4.2cm]{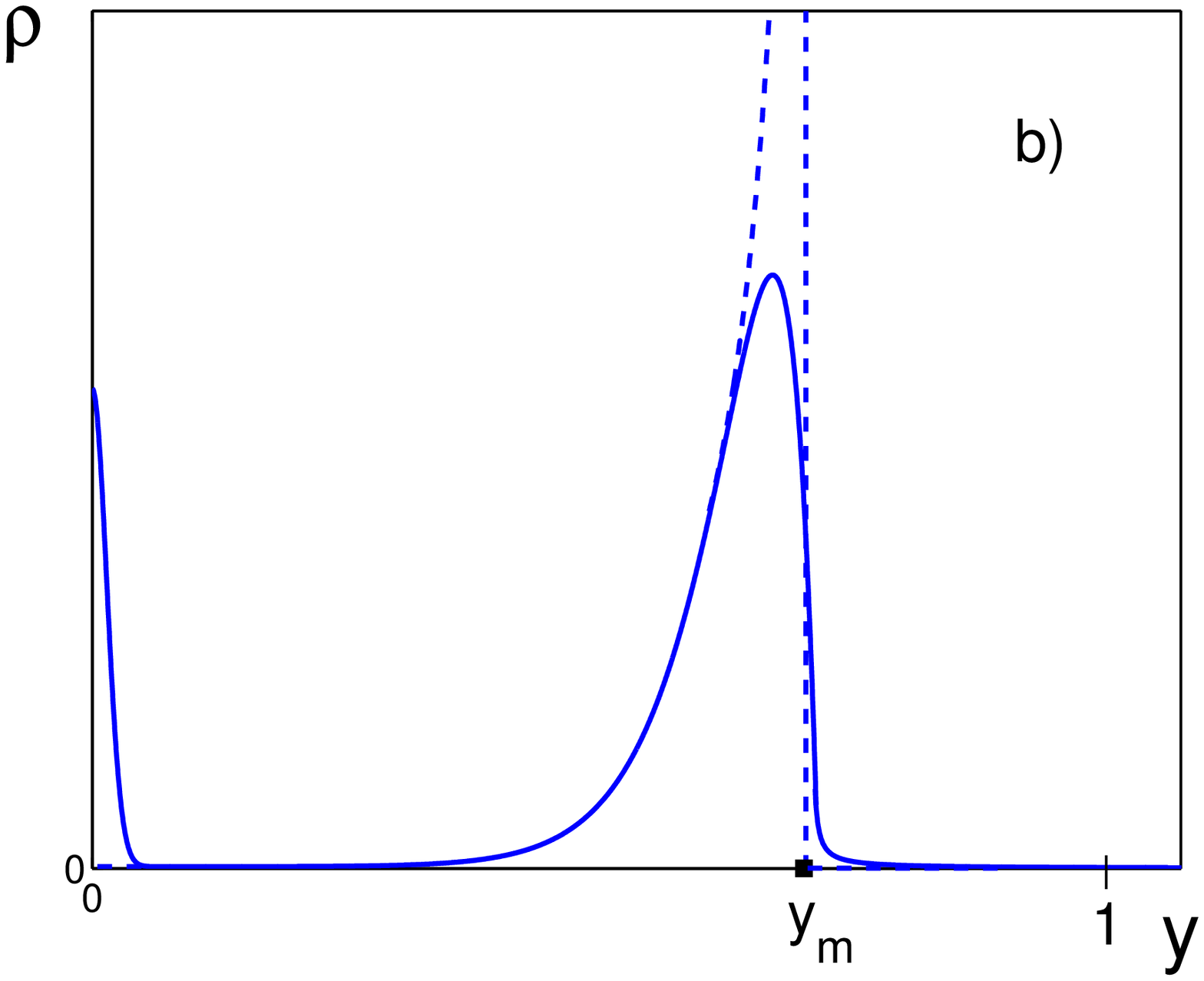}
   \includegraphics[height=4.2cm]{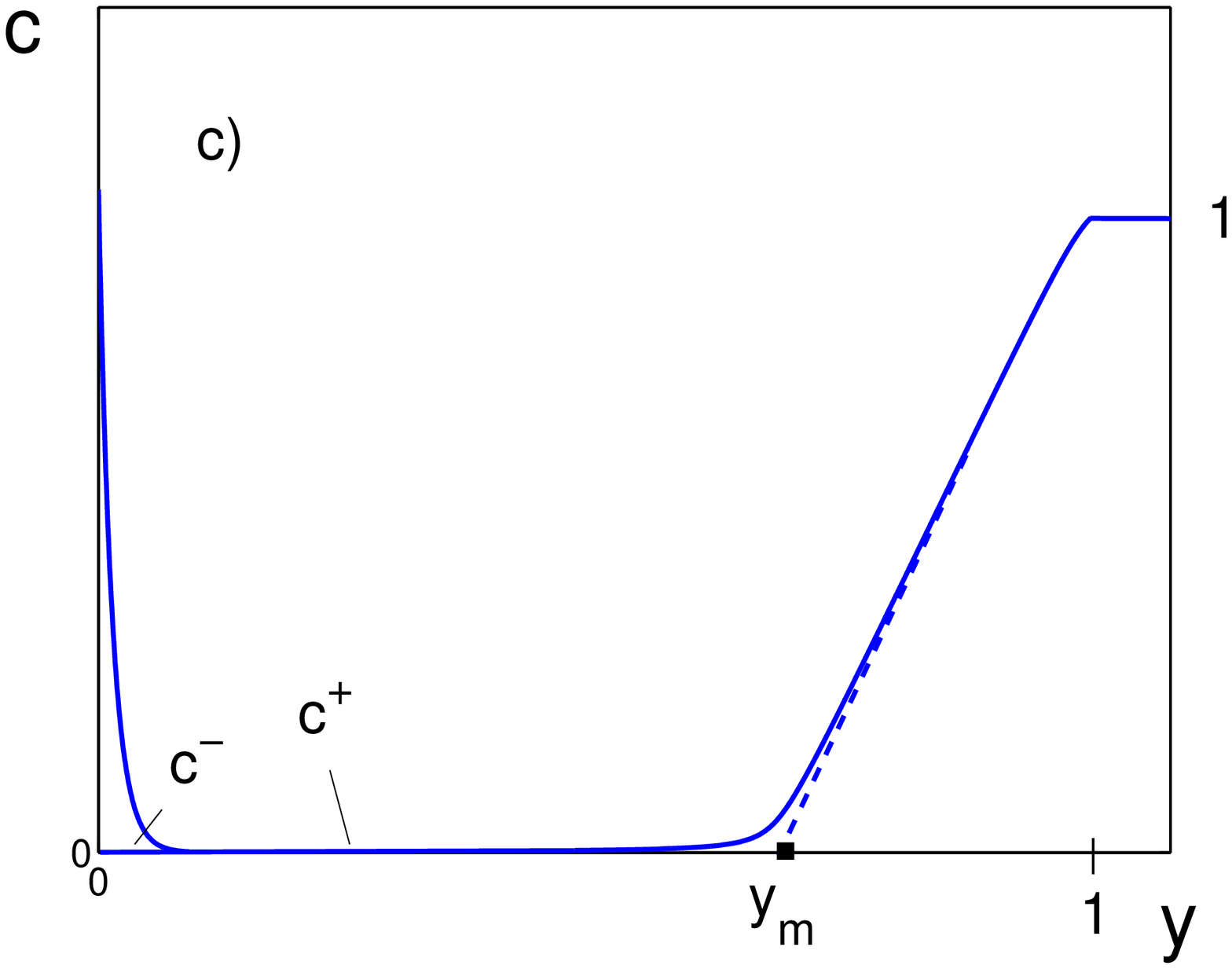}
\end{center}
\caption{Schematic for limiting processes in a steady--state
equilibrium: a)  electric field $E{\,=\,}d \Phi / dy$; b)
volume charge density $\rho{\,=\,c^+\,-\,}c^-$; c) ion
concentrations $c{\,=\,}c^+$ and $c{\,=\,}c^-$. The dashed line is
the outer solution  in the small vicinities of $y{\,=\,}0$ (the
double ion boundary layer) and $y{\,=\,}y_m$ (the boundary layer
between the volume space charge region and the diffusion layer).
The solid and dashed lines depict  exact and  asymptotic solutions
for small $\varepsilon$.}\label{P3}
\end{figure}

The solution \eqref{eq3020}--\eqref{eq3022} describes  extreme
non--equilibrium over--limiting regimes, while \eqref{eq3015} is
valid for the under--limiting regimes. A unified asymptotic
description of EDL valid from under--limiting to  extreme
non--equilibrium over--limiting regimes is developed by Zaltzman
and Rubinstein \cite{Rub13}.

\textit{Experimental evidence of the electro--convective
instability.} With further increase of the potential drop
between the membranes,   the over--limiting currents eventually
arise, and steady--state equilibria lose their applicability
(transition from the regime B to C in Fig.~\ref{P2}). In general,
four physical mechanisms can be responsible for that phenomenon.
Additional charge carriers  due to electrolyte splitting and
exaltation effects are assumed to be responsible for over--limiting
currents in Belova et al.~\cite{Belv}, Pismenskaya et
al.~\cite{Pismen}. Rayleigh--Benard convection   (see review by Cross and Hohenberg ~\cite{CrHog})
 can be also due to the Ohmic heating, hence, it can provide another physical mechanism responsible for the over--limiting regimes.
 However, for a relatively small
distance between membranes, the gravity--convection effect is
absent, since the Rayleigh number $Ra$ is less than the critical
value, $Ra_{cr}{\,=\,}1708$ (see experimental data in
~\cite{Belv}, \cite{Pismen}). Finally, Rubinstein, Staude and
Kedem \cite{Rub23} found experimentally that the transition  to
the overlimiting regime C (see Fig.~ \ref{P2}) is accompanied
 by current
oscillations which are regular at a small super--criticality and
irregular at a large super--criticality (regimes ~$C_I$ and
~$C_{II}$ in Fig.~\ref{P2}).
 The above--mentioned  facts  indirectly indicate to a
correlation between the instability and over--limiting currents.
The correlation between the electro--convective instability and the
over--limiting regimes is demonstrated in Maletzki et
al.~\cite{Mal} and Rubinstein et al.~\cite{Rub12}, where the
over--limiting regimes are eliminated when instability has been
artificially suppressed. The first direct experimental proof of
the electro--convective instability, which arises with increasing
  potential drop between the bulk ion--selective membranes, is
reported by Rubinstein et al.~\cite{RubRub1}, who manage to show the existence of small vortices near the membrane surface.   Yossifon and
Chang~\cite{YCh} also observe an array of electro--convective vortices
 arising under the effect of a slow AC electric field, while
Kim et al.~\cite{KmHn} observe  non--equilibrium electroosmotic
vortices.

\textit{Model of electroconvection as a stability lost of
steady--state equilibria.} As it was mentioned above, the
over--limiting currents are accompanied by an instability of the
membrane system. The following two modes of electroconvection in
strong electrolytes are distinguished: i) bulk electroconvection
due to the volume electric forces \cite{Grig}--\cite{LerRub}; and
ii)  common electroosmosis, either of the classical ``first''
kind \cite{Dukh}--\cite{Bst1} or of the ``second'' kind
\cite{Dukh}--\cite{DukhDe}.  In   the first kind electroosmosis, a
slip velocity results from the tangential electric field applied
to the space charge of a quasi--equilibrium EDL \cite{Rub13}, while
in the second kind  regimes, slip velocity results from the field
applied to  the ESC.

 The stability theory of 1D quiescent
steady--state  EDL is developed by Rubinstein and Zaltzman
\cite{Rub3},\cite{Rub3a} for the extreme nonequilibrium conditions
\eqref{eq3020}--\eqref{eq3022}. First, the  region
$0{\,\leqslant\,y\,\leqslant\,}L$ is subdivided into  the space
charge, $0{\,\leqslant\,y\, < \,}y_m$, and electro--neutral,
 $y_m{\,<\,y\,\leqslant\,}L$, regions. The latter consists of the
diffusion layer region, $y_m{\, <\,y\,<\,}1$, and the bulk neutral
region, $1{\,<\,y\,\leqslant\,}L$. For the over--limiting regimes,
the current inhomogeneity
 along the membrane leads to a
convective motion  of fluid with the tangential slip velocity,
$U_m$, which is determined from the solution in the ESC region,
$0<y \leqslant y_m$.  Then the   concentration, pressure and
velocity are determined in the electro--neutral region,
$y_m{\,<\,y\,\leqslant\,}L$, using the slip velocity found above
as the boundary value at $y=y_m$. The 1D steady--state equilibrium
solution is found to be stable for under--limiting regimes and
unstable for over--limiting regimes. The corresponding marginal
stability curve (in the plane  of the potential drop, $\Delta
\Phi$, and wave number, $\alpha$)  has no  upper branch, and
the growth rate
 increases infinitely with the wave number. A
regularized problem that takes into account the higher order terms
in $\varepsilon$ and predicts the upper branch,  is offered by
Rubinstein and Zaltzman \cite{Rub10}. Linear stability results for
various asymptotic
 formulations along with numerics are presented by
 Rubinstein, Zaltzman and Lerman \cite{RubLer}. Instability theory
 of 1D quiescent steady--state solution
uniformly valid for both equilibrium and non--equilibrium
conditions (not necessarily of the extreme type) is developed by
Zaltzman and  Rubinstein \cite{Rub13}.
 They obtain that non--equilibrium electroosmotic instability
occurs at the  transition from the quasi--equilibrium to the
non--equilibrium EDL regime.

Note the numerical simulations of the nonlinear unsteady problem
which demonstrate, in qualitative agreement with experimental
data, the formation of the pair vortices and their development to
chaotic motion in the electro--neutral region (\cite{Rub10, Pun2}).

\begin{figure}
  \begin{center}
    \includegraphics[height=7cm]{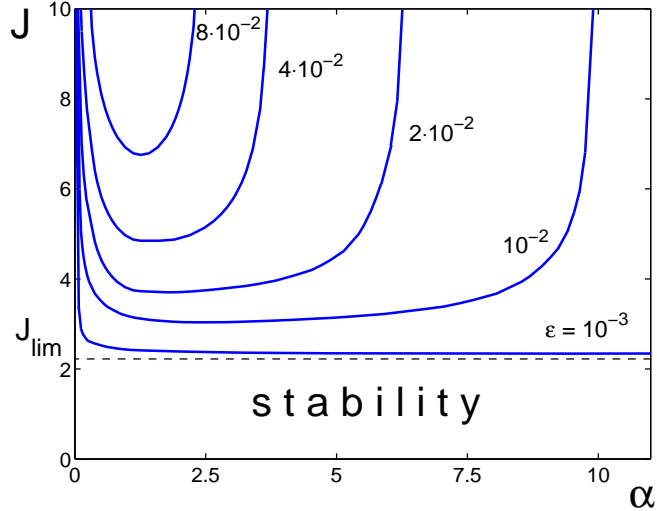}
  \end{center}
\caption{Marginal stability curves at different $\varepsilon$,
numerics. The under--limiting regimes are  stable (adopted from
Demekhin et al. \cite{Dem}).}\label{PAJ}
\end{figure}

\textit{New type of equilibria and their stability.} Although the
 cells observed in Rayleigh--Benard convection and
Benard--Marangoni thermo--convection look like the cells in the
electro--convective motion, the latter instability is  much more
complicated  from both physical and mathematical points of view. In
particular, electro--hydrodynamics of electrolytes between
membranes manifests new types of equilibria and their
instabilities  developed in micro-- and nano--scales.

The present numerical calculations evaluate the settling time
as about several minutes, while  experiments demonstrate that the
characteristic time  for the instability to manifest itself is
about several seconds (see \cite{YCh}, \cite{Belv}, \cite{Rub12}
and the Subsection IIC). Thus, although the problem solution eventually reaches a steady--state
equilibrium which becomes unstable, an unsteady solution can lose its
stability long before the relaxation to the steady state.
These two unperturbed solutions, steady--state and self--similar unsteady  ones,  are rather correspond to different, gradual and instantaneous  loading of the potential drop to the membrane system, respectively.

The existence of this novel type of  unsteady solutions has been
recently predicted numerically for small $\varepsilon$ at
intermediately long times \cite{Dem} and observed experimentally
\cite{YCh}. In the present work, unsteady transient equilibria and
their stability are investigated both asymptotically and
numerically. The present asymptotic analysis is restricted by the
limiting regimes and their transition to the over--limiting
regimes. The results of the present asymptotic analysis are
consistent with the exact numerical solution.

 The paper
is structured as follows. In the next Section,
the governing relations  are presented and typical characteristic values of the system parameters are evaluated.
The 1D unsteady  and 1D self--similar problems are developed for intermediately long times.
Numerical justification of closeness of the exact  solution of the initial--boundary problem to the solution of the self--similar problem is justified numerically  for intermediately long times.
 In  Section III, 1D explicit  solutions of the self--similar problem are obtained for small Debye length and
compared with the numeric  solutions for intermediately long times.
A two--parametric family of self--similar unsteady equilibrium solutions is found and
their stability is investigated. The justification of
self--similarity  is done both
numerically and asymptotically. In Section IV, the unsteady   2D problem  is made dimensionless,  and  simplified  equations
are derived for small Debye length.
Section V describes the linear stability  of the 1D self--similar
solutions slowly varying with time.  The resulting marginal
stability curves  are compared with those obtained by direct
numerical simulations.
Some of cumbersome calculations for electrostatic solution in the space charge region and for slip velocity are in Appendixes A and B, respectively.

{\color{black}
\section{THE PHYSICAL MODEL AND \\1D SELF--SIMILAR PROBLEM}
\subsection{Governing equations and typical values of parameters}
Electro--convection in a binary electrolyte between semi--selective
ion--exchange  membranes is described by the equations for  ion
transport, the Poisson equation for the electric potential and
the Stokes equation for a creeping  flow (this model was first
introduced in \cite{Rub13}):
\begin{equation}\label{eq1Dim}
 \frac{\partial \tilde{c}^\pm}{\partial \tilde{t}}+\tilde{\textbf{U}}\cdot\tilde{\nabla}\tilde{c}^\pm = \tilde{D}\left\{ \  \pm\frac{\tilde{F}}{\tilde{R}\tilde{T}}\tilde{\nabla}\cdot\left(\tilde{c}^\pm\tilde{\nabla}\tilde{\Phi}\right)+\tilde{\nabla}^2\tilde{c}^\pm\right\},
\end{equation}
\begin{equation}\label{eq112Dim}
 \tilde{\nabla}^2\tilde{\Phi}=\frac{\tilde{F}}{\tilde{\epsilon}}\left(\tilde{c}^--\tilde{c}^+\right),
\end{equation}
\begin{equation}\label{eq113Dim}
-\tilde{\nabla}
\tilde{P}+\tilde{\mu}\tilde{\nabla}^2\tilde{\textbf{U}}+\tilde{F}\left(\tilde{c}^--\tilde{c}^+\right)\tilde{\nabla}\tilde{\Phi}=0, \
\ \ \ \tilde{\nabla}\cdot\tilde{\textbf{U}}=0,
\end{equation}
This system of  dimensional equations is complemented by the
proper initial and boundary conditions which are as follows
\begin{equation}\label{eq20007Dim}
\tilde{c}^+ =\tilde{p},\ \ \ \displaystyle -\frac{\tilde{F}\tilde{c}^-}{\tilde{R}\tilde{T}}\frac{\partial \tilde{\Phi}}{\partial \tilde{y}}+\frac{\partial \tilde{c}^-}{\partial \tilde{y}}=0,\ \ \ \tilde{\Phi} =0,\ \ \ \tilde{\textbf{U}} =0\ \ \ \mbox{for}\ \ \ \tilde{y}=0,
\end{equation}
\begin{equation}\label{eq20002Dim}
\tilde{c}^+ =\tilde{p},\ \ \ \displaystyle -\frac{\tilde{F}\tilde{c}^-}{\tilde{R}\tilde{T}}\frac{\partial \tilde{\Phi}}{\partial
\tilde{y}}+\frac{\partial \tilde{c}^-}{\partial \tilde{y}}=0, \ \ \ \tilde{\Phi} =\Delta\tilde{V}, \ \ \ \tilde{\textbf{U}}=0\ \ \ \mbox{for}\ \ \ \tilde{y}=L,
\end{equation}
\begin{equation}\label{eq277Dim}
\tilde{c}^+ = \tilde{c}^- =\tilde{c}_{\infty}\ \ \ \mbox{for}\ \ \ \tilde{t}=0.
\end{equation}
Here $\tilde{c}^+,\ \tilde{c}^-$ are the molar concentrations of
cations and anions;
$\tilde{\textbf{U}}=\{\tilde{U},\:\tilde{V},\:\tilde{W}\}$ is the
fluid velocity; $\{\tilde{x},\:\tilde{y},\:\tilde{z}\}$ are  the
coordinates,  $\tilde{y}$ is normal  to the membrane surface;
$\tilde{\Phi}$ is the electrical potential; $\tilde{\epsilon}$
is the permittivity of the medium; $\tilde{P}$ is the pressure;
$\tilde{c}_{\infty}$ is the initial ion concentration; $\tilde{F}$
is Faraday's constant; $\tilde{R}$ is the universal gas constant;
$\tilde{T}$ is an  absolute temperature; $\tilde{\mu}$ is the
dynamic viscosity of the fluid; $\tilde{D}$ is the cationic and
anionic diffusivity; $\tilde{p}$ is  the interface concentration.

It is  convenient to use the characteristic electric current
$\tilde{j}$ at $\tilde{y}=0$  which is related with the known
total  potential  drop  between membranes $\Delta\tilde{ V}$
\begin{equation}\label{eqJPDim}
\displaystyle\tilde{j} = \frac{\tilde{D}\tilde{F^2}
}{\tilde{R}\tilde{T}}\,\tilde{c}^+\frac{\partial\tilde{\Phi}}
{\partial\tilde{y}}+\tilde{D}\tilde{F}\frac{\partial\tilde{c}^+}{\partial\tilde{y}},\
\ \ \ \ \ \tilde{y} = 0.
\end{equation}

Let us evaluate  the  parameters of the system. The  bulk
concentration of the aqueous  electrolytes varies in the range
 $\tilde{c}_{\infty}{\,=\,1\,\div\,10^{3}\ }mol/m^3$; the potential
drop is about $\Delta\tilde{V}{\,=\,0\,\div\,5\ }V$; the  absolute
temperature can be taken as $\tilde{T}{\,=\,300^0\ }K$; the
diffusivity is about $\tilde{D}{\,=\,2\,\cdot10^{-9}\ }m^2/s$; the
 distance between the electrodes $\tilde{L}$ is of the
order of $0.5{\,\div\,}1.5$ $mm$; the concentration value
$\tilde{p}$ on the membrane surface must be much higher than
$\tilde{c}_{\infty}$, and it is usually taken within the range
from $5\tilde{c}_{\infty}$ to $10\tilde{c}_{\infty}$
 ( \cite{Rub1}, \cite{Rub13}, \cite{Zab}). The dimensional Debye layer thickness
$\tilde{\lambda}_D$ is varied in the range from $0.5$ to $15 \
nm$ depending on the concentration $\tilde{c}_{\infty}$.

\subsection{1D unsteady problem  at intermediately long times}
 Let us consider an 1D unsteady equilibrium state:
\begin{equation}\label{equilibr}
 \tilde{\textbf{U}}= 0,\qquad \frac{\partial }{\partial \tilde{x}}=\frac{\partial }{\partial \tilde{z}}=0.
\end{equation}
Then the system \eqref{eq1Dim}--\eqref{eqJPDim} turns into
\begin{equation}\label{eqN111}
      \frac{\partial \tilde{c}^\pm}{\partial \tilde{t}}=
      \tilde{D}\left\{\pm \frac{\tilde{F}}{\tilde{R} \tilde{T}}
      \frac{\partial}{\partial \tilde{y}}\left(
      \tilde{c}^\pm\frac{\partial\tilde{\Phi}}{\partial \tilde{y}}\right)+
      \frac{\partial^2 \tilde{c}^\pm}{\partial \tilde{y}^2}\right\},
\end{equation}
\begin{equation}\label{eqN131}
      \frac{\partial^2 \tilde{\Phi}}{\partial \tilde{y}^2}=
      \frac{\tilde{F}}{\tilde{\epsilon}}(\tilde{c}^--\tilde{c}^+),
\end{equation}
\begin{equation}\label{eq2000111}
\tilde{c}^+ =\tilde{p},\ \ \ \displaystyle -\frac{\tilde{F}\tilde{c}^-}{\tilde{R} \tilde{T}} \frac{\partial
\tilde{\Phi}}{\partial \tilde{y}}+\frac{\partial\tilde{c}^-}{\partial \tilde{y}}=0, \ \ \ \ \ \tilde{\Phi} = 0\ \ \ \mbox{for}\ \ \ \tilde{y}=0,
\end{equation}
\begin{equation}\label{eq2000121}
\tilde{c}^+ =\tilde{p},\ \ \ \displaystyle
-\frac{\tilde{F}\tilde{c}^-}{\tilde{R} \tilde{T}} \frac{\partial\tilde{\Phi}}{\partial \tilde{y}}+\frac{\partial\tilde{c}^-}{\partial \tilde{y}}=0, \ \ \ \ \ \tilde{\Phi}=\Delta\tilde{V}\ \ \ \mbox{for}\ \ \
\tilde{y}=\tilde{L},
\end{equation}
\begin{equation}\label{eq2770P11}
\tilde{c}^+ =\tilde{c}^- =\tilde{c}_{\infty}\ \ \ \mbox{for}\ \ \ \tilde{t} =0,
\end{equation}
\begin{equation}\label{eqJPDim11}
\displaystyle\tilde{j} = \frac{\tilde{D}\tilde{F^2}
}{\tilde{R}\tilde{T}}\,\tilde{c}^+\frac{\partial\tilde{\Phi}}
{\partial\tilde{y}}+\tilde{D}\tilde{F}\frac{\partial\tilde{c}^+}{\partial\tilde{y}},\ \ \ \tilde{y} = 0.
\end{equation}
 According to  \eqref{eqN111}--\eqref{eqJPDim11}, at $\Delta\tilde{V}=0$, a neutral steady--state
 solution with a uniform concentration of  ions, $ \tilde{c}^{+}{\,=\,\tilde{c}^{-}=\,}\tilde{c}_{\infty}$, instantaneously
 loses its equilibrium under the effect of  the potential
 drop, $\Delta\tilde{V}$, applied at
 $\tilde{t}{\,=\,}0$  and kept constant at $\tilde{t}{\,>\,}0$.
As a result, a concentration polarization of ions is formed in
thin boundary layers near  the membrane surfaces, which expand with
time. After several milliseconds of evolution, the diffusion layer
thickness, $\tilde{\delta}$, will be much larger than the Debye
length,
$\tilde{\lambda}_D{\,=\,}(\tilde{\epsilon}\tilde{R}\tilde{T}/\tilde{c}_\infty)^{1/2}/\tilde{F}$,
$\quad \tilde{\delta}(\tilde{t}) \gg \tilde{\lambda}_D. $ On the
other hand, there is a wide interval of time $\tilde{t}$ (from
seconds to minutes),  such that $ \tilde{\delta}(\tilde{t}) \ll
\tilde{L}$:
\begin{equation}\label{EstimatDifThick}
\tilde{\lambda}_D\ll \tilde{\delta}(\tilde{t}) \ll \tilde{L}.
\end{equation}
We assume that the diffusion--layer thickness
$\tilde{\delta}(\tilde{t})$ in the 1D unsteady solution  varies in
time as
\begin{equation}\label{delta12}
\tilde{\delta}(\tilde{t})=2\sqrt{\tilde{D}\tilde{t}}.
\end{equation}
Then \eqref{EstimatDifThick} can be  presented in the form
\begin{equation}\label{timeeva1}
\frac{\tilde{\lambda}_D^2}{4 \tilde{D}} \ll \tilde{t} \ll
\frac{\tilde{L}^2}{4 \tilde{D}}.
\end{equation}

Let us study the problem for intermediately long times which
belong to the interval \eqref{timeeva1}. The spatial structure of
the unsteady limiting regimes  at any fixed time is qualitatively
similar to that of the steady--state equilibria depicted in
Fig.~\ref{P3}. The space between the membranes  can be
 divided into the following regions of inhomogeneity: (I)
a thin EDL  near the membrane surface
 $\tilde{y}{\,=\,}0$;  ESC region   (II)  separated from  the diffusion layer (III) by
 a thin internal boundary layer (IV);   a bulk neutral region (V).

For unsteady solutions, the influence of the upper membrane
$\tilde{y}{\,=\,}\tilde{L}$ is neglegible within the interval
\eqref{timeeva1}, and  the boundary conditions \eqref{eq2000121}
should be revised. Subtracting the second equation \eqref{eqN111}
for $\tilde{c}^-$ from the first equation \eqref{eqN111} for
$\tilde{c}^+$, integrating the result from $0$ to some $\tilde{y}$
within the bulk neutral region, and neglecting the higher order
terms with respect to small $\tilde{\delta}/\tilde{L}$, the
known linear voltage distribution in the bulk neutral region
 is obtained
\begin{equation}\label{PhiJ}
 \tilde{\Phi}(\tilde{y}) \sim const
 +\frac{\tilde{j}\tilde{R}\tilde{T}}{2\tilde{F}^2\tilde{D}\tilde{c}_\infty}\,\tilde{y},\ \ \ \tilde{y}\gg\tilde{\delta}.
 \end{equation}
Here the $constant$ and the electric current $\tilde{j}$ are
functions of time. The relation \eqref{PhiJ} predicts
unbounded linear behavior of the potential at a large distance from
the bottom membrane. This infers the following effective boundary
condition at large $y$
\begin{equation}\label{PhiJ1}
\lim_{\tilde{y}/\tilde{\delta}\rightarrow\infty}\left(\tilde{\Phi}- \frac{\tilde{j}\tilde{R}\tilde{T}}{2\tilde{F}^2\tilde{D}\tilde{c}_\infty}\,\tilde{y}\right)=\Delta\tilde{\Phi},
\end{equation}
which, along with the following condition for the concentrations
\begin{equation}\label{eq33}
\lim_{\tilde{y}/\tilde{\delta}\rightarrow\infty}\tilde{c}^+ =\lim_{\tilde{y}/\tilde{\delta}\rightarrow\infty}\tilde{c}^- =\tilde{c}_\infty
\end{equation}
substitute the BC's \eqref{eq2000121} for the problem with the
distance between the upper and bottom membranes that satisfies \eqref{EstimatDifThick}.
Note that Ohmic current is taken into account due to a cumulative
effect at a large distance from the membrane.

Introducing a similar effective voltage instead of the
total voltage $\Delta V$ is useful also for the finite membrane
problem both for the efficiency of calculations and comparison with
the case of the distant upper membrane:
\begin{equation}\label{PhiJ2}
\Delta \tilde{\Phi} = \Delta \tilde{V}-\frac{\tilde{j}\tilde{R}\tilde{T}\tilde{L}}{2\tilde{F}^2\tilde{D}\tilde{c}_\infty}.
\end{equation}
In fact, the effective voltage $\Delta \tilde{\Phi}$ in the EDL and ESC
regions does not depend on the distance between the
membranes.

The formulated problem does not contain a characteristic length,
since neither the distance between the electrodes nor the double--ion
length can give such a scale within the interval \eqref{timeeva1}.
This infers the solution  self--similarity \cite{Sed} with the
diffusion layer thickness as the dynamic characteristic size
( justification of self--similarity  in
small $\varepsilon$ see in Section II.C). Let us
transform the problem into a dimensionless form using the
characteristic values of potential,
$\displaystyle\tilde{\Phi}_0{\,=\,}\tilde{R}\tilde{T}/\tilde{F}$,
 concentration, $\tilde{c}_\infty$, and  time, $\tilde{t}_0$, by introducing the following new
independent variables:
\begin{equation}\label{eq33dop}
\displaystyle{\tau}={t},\ \ \ \
\eta=\frac{\tilde{y}}{\tilde{\delta}(\tilde{t})}.\,\,
\end{equation}
This gives
\begin{equation}\label{eq141}
\begin{array}{c}
\displaystyle
      \frac{\partial c^+}{\partial \tau}=2\eta\frac{\partial
      c^+}{\partial\eta}+
      \frac{\partial}{\partial\eta}\left(
      c^+\frac{\partial\Phi}{\partial\eta}\right)+
      \frac{\partial^2c^+}{\partial\eta^2},\\[14pt]
\displaystyle
     \frac{\partial c^-}{\partial \tau}=2\eta\frac{\partial c^-}{\partial\eta}-
      \frac{\partial}{\partial\eta}\left(
      c^-\frac{\partial\Phi}{\partial\eta}\right)+
      \frac{\partial^2c^-}{\partial\eta^2},\\[14pt]
\displaystyle
      \varepsilon^2\frac{\partial^2\Phi}{\partial\eta^2}=c^--c^+,\\
\end{array}
\end{equation}
where $\displaystyle
\varepsilon{\,=\,}\tilde{\lambda}_D/\tilde{\delta}(\tilde{t})$.

Assume now that for intermediately long times the solution,
indeed, becomes self--similar (see Section II.C), i. e.
\begin{equation}\label{eq41}
\begin{array}{c}
\displaystyle 2\eta\frac{d c^+}{d\eta}+\frac{d}{d\eta}
\left(c^+E+\frac{d c^+}{d\eta}\right)=0,\\[14pt]
\displaystyle 2\eta\frac{d c^-}{d\eta}+\frac{d}{d\eta}
\left(-c^-E+\frac{d c^-}{d\eta}\right)=0,\\[14pt]
\displaystyle \varepsilon^2\frac{d E}{d\eta}=c^--c^+,
\end{array}
\end{equation}
with the BC's
\begin{equation}\label{eq42}
\begin{array}{cc}
 & \ \ \ c^+=p,\ \ \ \displaystyle c^-E-\frac{d c^-}{d\eta}=0,\ \ \
\Phi = 0\ \ \ \mbox{for}\ \ \ \eta=0,
\end{array}
\end{equation}
\begin{equation}\label{eq43}
\begin{array}{cc}
 & \ \ \ \displaystyle c^-\to1,\ \ \ c^+\to1,\ \ \ \displaystyle\Phi-\frac{J\eta}{2}\to\Delta\Phi\ \ \ \mbox{at}\ \ \ \eta\to\infty.
\end{array}
\end{equation}
Here $\displaystyle E{=\,}\frac{d \Phi}{d\eta}$, $\Delta\Phi{\,=\,}\Delta\tilde\Phi/\tilde\Phi_0$, the dimensionless
electric current at $\eta=0$ is determined by
\begin{equation}\label{eq44}
J=c^+E+\frac{d c^+}{d\eta},\ \ \ \eta=0.
\end{equation}

Solutions of the self--similar boundary problem \eqref{eq41}--\eqref{eq44} are presented in Fig.~\ref{rho} for the charge
distribution $\rho(\eta){\,=\,c^+(\eta)\,-\,}c^-(\eta)$ and
positive ions concentration $c^+(\eta)$ for different values of
the parameters.  Voltage--current characteristics $J$ versus $\Delta
\Phi$ for different  $\varepsilon$  are shown in
Fig.~\ref{VACM} for both the exact numeric and   asymptotic
self--similar solutions.

\begin{figure}
\begin{center}
    \includegraphics[width=7.85cm]{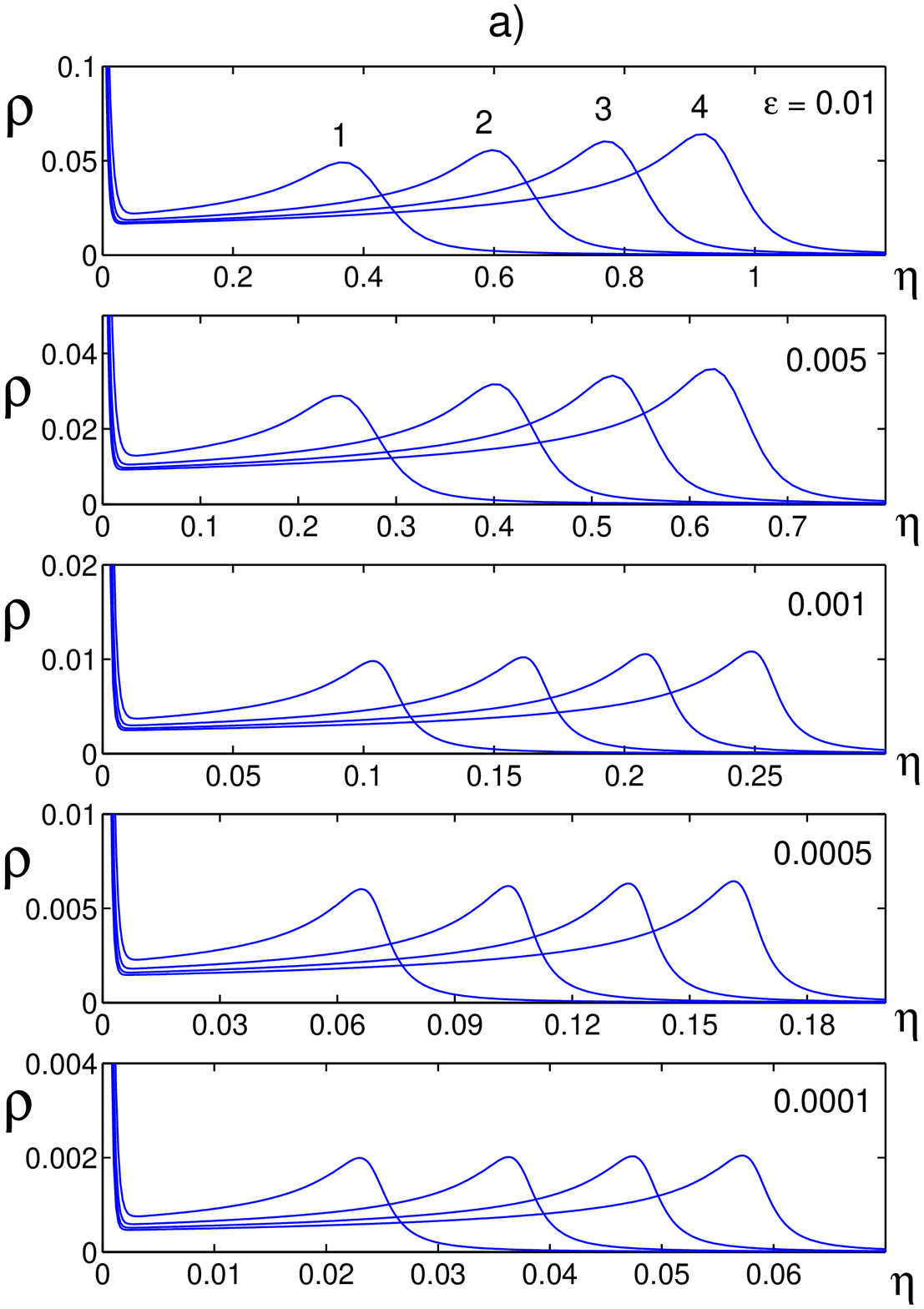}
    \includegraphics[width=7.85cm]{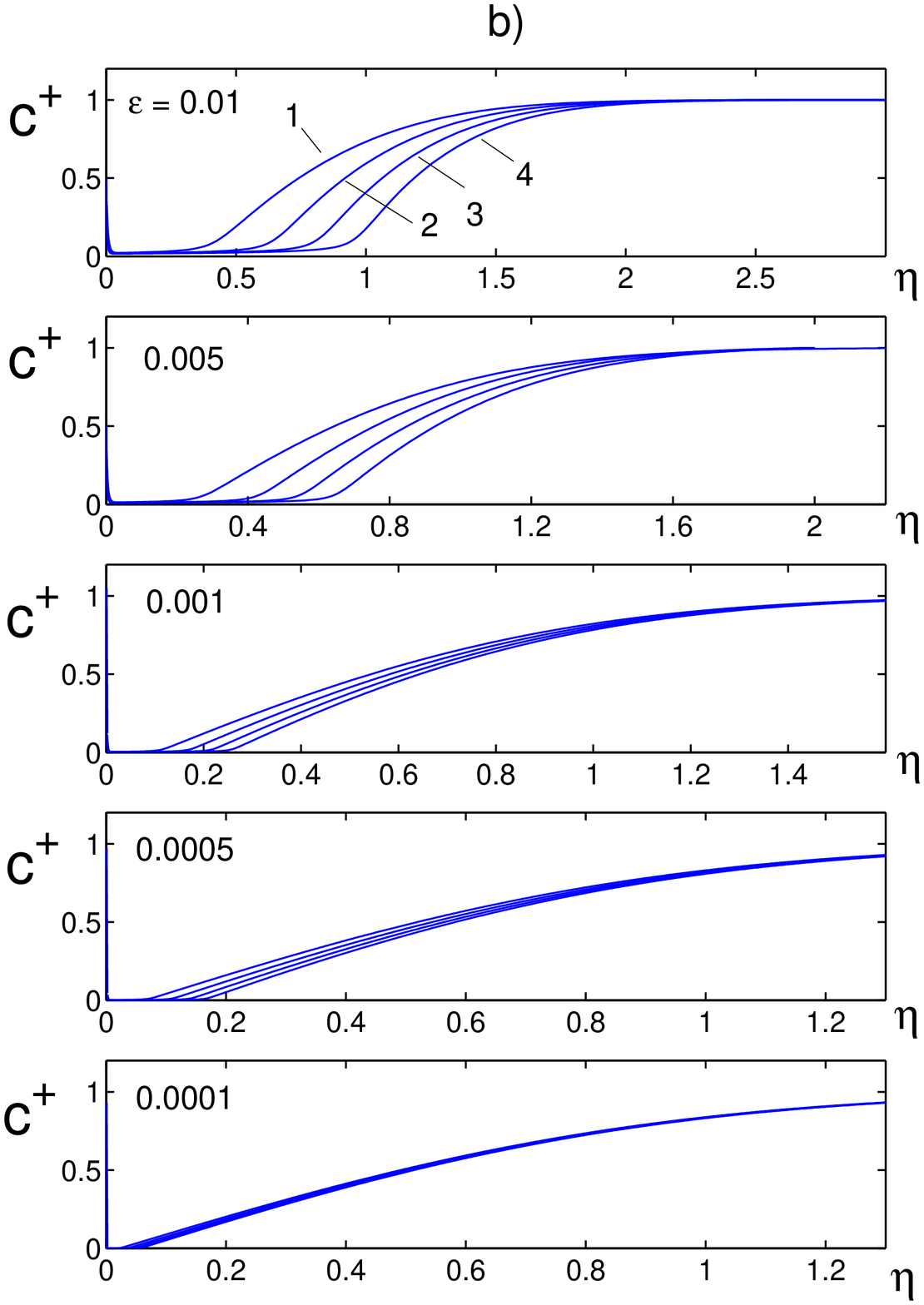}
\end{center}
\caption{Self--similar charge $\rho=c^+-c^-$ (a) and concentration
$c^+$ distribution (b) for different $\varepsilon$. Curves
correspond to the following values of electric potential drop: 1.
$\Delta \Phi=50$; 2. $\Delta \Phi=100$; 3. $\Delta \Phi=150$; 4.
$\Delta \Phi=200$.}\label{rho}

\end{figure}
\begin{figure}
\begin{center}
    \includegraphics[height=8cm]{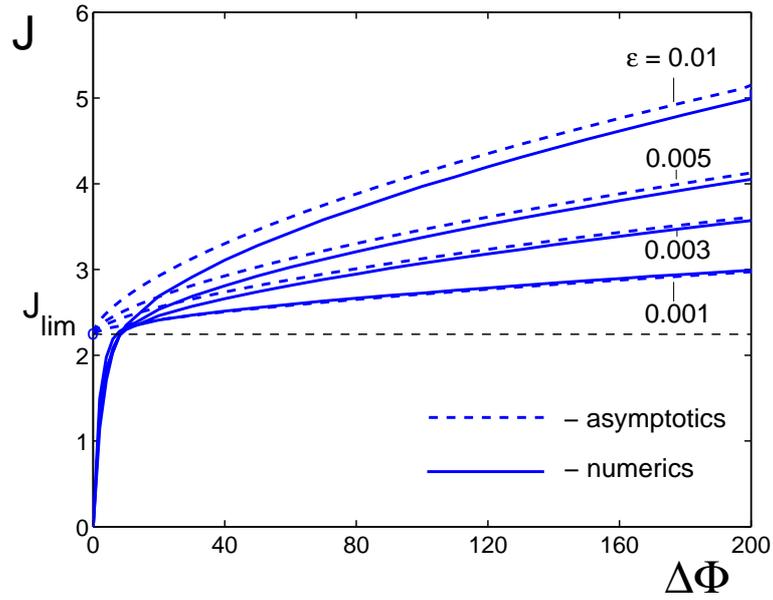}
\end{center}
\caption{VC curves. Comparison of the self--similar asymptotics
(dashed lines) and the exact numerics (solid lines) for several
$\varepsilon$.}\label{VACM}
\end{figure}

\begin{figure}
\begin{center}
    \includegraphics[height=8cm]{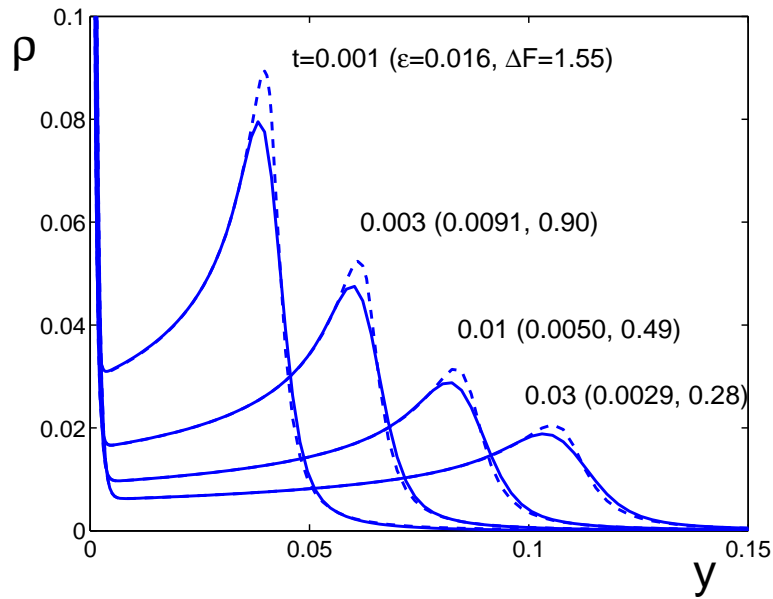}
\end{center}
\caption{Distribution of the charge density $\rho$ in space for
several moments of time, $\Delta V=100$ and $\nu{\,=\,}\tilde{\lambda}_D/\tilde{L}=0.001$. Solid
line corresponds to the numerical solution of the unsteady
problem \eqref{eqN1111}--\eqref{eq2770P111}. Two--parametric
self--similar solution of \eqref{eq41}--\eqref{eq44} with
$\varepsilon$ and $\Delta \Phi$ parametrically changing with time is
depicted by dashed lines.}\label{rho3}
\end{figure}

\subsection{Numerical justification of the self--similar problem}

To justify self--similarity of the solution of \eqref{eq41}--\eqref{eq44} for intermediately long times,
we compare it with the exact numerical solution of the initially--boundary problem
\eqref{eqN111}--\eqref{eqJPDim11}. In this numeric simulations, the time
$\tilde t$ and the coordinate $\tilde{y}$ are referred to
$\tilde{L}^2/\tilde{D}$ and $\tilde{L}$. The concentrations
$\tilde{c}^{\pm}$ and the potential $\tilde{\Phi}$ are referred to
$\tilde{c}_{\infty}$ and $\tilde{\Phi}_{0}$, respectively, as it
was done for the self--similar solution. The system
\eqref{eqN111}--\eqref{eqJPDim11} turns into
\begin{equation}\label{eqN1111}
      \frac{\partial c^\pm}{\partial t}=
      \pm \frac{\partial}{\partial y}\left(
      c^\pm\frac{\partial \Phi}{\partial y}\right)+
      \frac{\partial^2 c^\pm}{\partial y^2},
\end{equation}
\begin{equation}\label{eqN1311}
    \nu^2  \frac{\partial^2 \Phi}{\partial y^2}=c^--c^+,
\end{equation}
\begin{equation}\label{eq20001111}
c^+ =p,\ \ \ \displaystyle -c^- \frac{\partial \Phi}{\partial
y}+\frac{\partial c^-}{\partial y}=0,\ \ \ \Phi = 0\ \ \ \mbox{for}\ \ \ y=0,
\end{equation}
\begin{equation}\label{eq20001211}
c^+ =p,\ \ \ \displaystyle -c^- \frac{\partial \Phi}{\partial
y}+\frac{\partial c^-}{\partial y}=0,\ \ \ \Phi =\Delta V \ \ \ \mbox{for}\ \ \ y=1,
\end{equation}
\begin{equation}\label{eq2770P111}
c^+ =c^- =1,\ \ \ \displaystyle\mbox{for}\ \ \ t =0,
\end{equation}
\begin{equation}\label{eqJPDim111}
\displaystyle j = \,c^+\frac{\partial \Phi} {\partial
y}+\frac{\partial c^+}{\partial y},\ \ \ y = 0,
\end{equation}
where $\nu{\,=\,}\tilde{\lambda}_D/\tilde{L}$, $\Delta V{\,=\,}\Delta\tilde V/\tilde\Phi_0$, $j{\,=\,}\tilde j\tilde L/(\tilde D\tilde F\tilde c_\infty)$.

The system \eqref{eqN1111}--\eqref{eqJPDim111} is integrated numerically
using Gear's method in time along  with Galerkin's $\tau$--method
digitization with respect to the space variable. The problem is
described by two parameters, the voltage between membranes,
$\Delta V$, and analog of $\varepsilon$ in a finite size
system, $\nu \ll 1$. The
calculations are fulfilled for the limiting regimes, $\nu$ and
$\Delta V$ are varied in the ranges $\nu{\,=\,}0.00001\div0.01$ and
$\Delta V{\,=\,}50\div200$. It is found numerically that the
solution reaches the steady--state equilibrium at
$t_s{\,=\,0.1\,\div\,}0.2$. Taking characteristic values
$\tilde{D}{\,=\,2\,\cdot10^{-9}\ }m^2/s$ and
  $\tilde{L}=$ $0.5{\,\div\,}1.5$ $mm$, this time is evaluated in a dimensional
  form as $\tilde{t}_s=$ $0.25{\,\div\,}4$ $min$, supporting the assumption in Subsection AI.
Below, we restrict ourselves by the regimes  satisfying the
inequality \eqref{timeeva1} which can be generalized now as
\begin{equation}\label{timeeva2}
 \frac{\nu^2}{4} \ll t \ll t_s.
\end{equation}

 The two--parametric family of  self--similar solutions of
the problem \eqref{eq41}--\eqref{eq44} is calculated for a wide
range of the parameters $\Delta \Phi$ and $\varepsilon$ and
accumulated for further comparison with the exact numerical solution
for the finite--length membrane geometry.  For the comparison of the
self--similar solution  with  the exact numerical solution for the
finite--length membrane geometry, the  parameters
$\Delta \Phi$ and $\varepsilon$ varying with time  are calculated  using relations
\begin{equation}\label{num100}
 \varepsilon(t)=\frac{\nu}{2 \sqrt{t}},\quad J(t)=2j(t)\sqrt{t}, \quad\Delta \Phi(t)=\Delta
 V-\frac{j(t)}{2}.
\end{equation}
Here the Ohmic voltage, $j(t)/2$, is excluded from the total
voltage $\Delta V$. These values are substituted into the accumulated
family of  self--similar solutions \eqref{eq41}--\eqref{eq44}
obtained previously for a wide range of the two values $\Delta
\Phi$ and $\varepsilon$.

The  charge density $\rho{\,=\,c^+\,-\,}c^-$ is presented in
Fig.~\ref{rho3} for both exact and approximate self--similar
solutions. The unsteady solution is ``adiabatically" sliding along
the two--parametric self--similar solution, $\Delta \Phi(t)$ and
$\varepsilon(t)$. One can see a rather good correspondence between
these two solutions for intermediately long times.

 This
expands the applicability  of the approach  up to  quite realistic
values of the potential drop. Thus, the present asymptotic  model
is applied to the dimensionless  potential drop $0 < \Delta
\Phi < 200 $ (which corresponds to dimensional values $0 \div 5V$).
The asymptotic model describes the limiting regimes with the error
value depending on $\varepsilon$. For $\varepsilon=0.01$ and
$\Delta \Phi=40$ the error is about $20\%$, while for $\Delta
\Phi=200$ and the same $\varepsilon=0.01$ it is about $3\%$. For
$\varepsilon=0.001$ the error is much smaller, for example for
$\Delta \Phi=20$ it is about $1\%$ and for $\Delta \Phi=200$ less
than $0.2\%$.

For the self--similar solution the diffusion layer thickness should
be inversely proportional to the electric current, $\delta(t) \sim
1/j(t) \sim \sqrt{t}$.
Typical evolution of $1/j(t)$ presented in Fig.~\ref{invJ}
exhibits three characteristic time intervals. In the interval
$II$, for the intermediately long times \eqref{timeeva2}, $1/j(t)$ is
proportional to $\sqrt{t}$ and the assumption \eqref{delta12} is
satisfied, while for  short times, in the time interval $I$,
as well as for long times, in the interval $III$, the
self--similarity is violated.

\begin{figure}
\begin{center}
    \includegraphics[height=8cm]{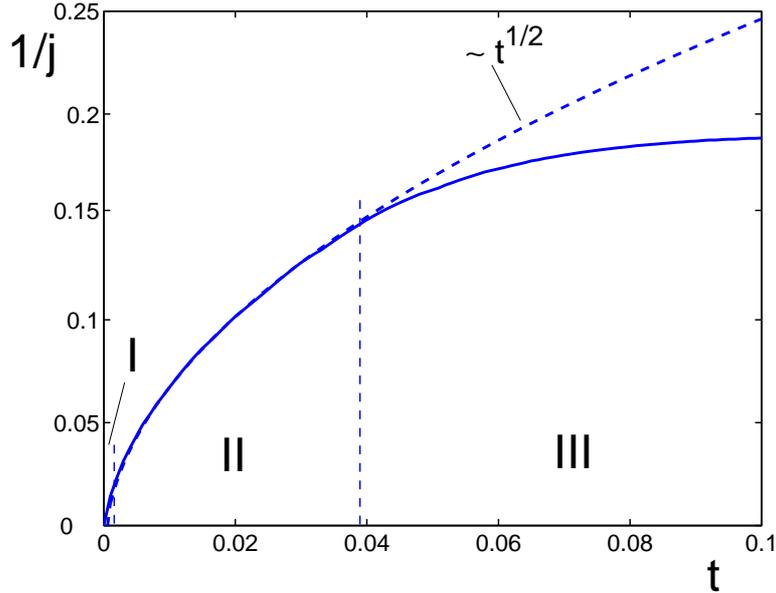}
\end{center}
\caption{Typical evolution of $1/j(t)$. I is the short--time region
of influence of the initial data; II is the intermediately--long--time
region of self--similarity, and III is the region of influence of
the upper membrane $y{\,=\,}1$, ($\Delta V{\,=\,}100$ and
$\nu{\,=\,}\tilde{\lambda}_D/\tilde{L}{\,=\,}0.001$).}\label{invJ}
\end{figure}

\section{1D SELF--SIMILAR SOLUTIONS FOR\\SMALL DIMENSIONLESS DEBYE LENGTH }
\subsection{1D explicit  solutions }
To find  1D self--similar equilibrium solutions explicitly, we apply to \eqref{eq41}--\eqref{eq44} the asymptotic decomposition
method widely used for 1D steady--state equilibria at small
$\varepsilon$, \cite{UrtKir}--\cite{Urt3}. The  reduced problem
allows a relatively simple solution for the electric field, which
consists of two branches on the left and right of the boundary
of the  space--charge region $\eta{\,=\,}\eta_m$. The value $\eta_m$ is
determined by the solution of the  problem. The resulting outer
solution is weakly singular at $\eta{\,=\,}\eta_m$, where the
electric field gradient has a discontinuity. This procedure
ignores
 the EDL near the membrane surface $\eta{\,=\,}0$ (the first boundary
 condition $c^+=p$ in \eqref{eq42} is rejected), but
 predicts the existence and location of the internal
 boundary layer in the vicinity of $\eta{\,=\,}\eta_m$. The influence of these boundary layers
on the limiting regimes is negligibly small  for vanishing
$\varepsilon$. In particular, the potential drop within these
boundary layers is much less by the order in $\varepsilon$ than
that within the ESC region. As in the steady--state case, the
matching conditions of the outer solution with the inner ones
within the boundary layers are  substituted by the patching
conditions. Such a description, in spite of its simplicity, well
describes solutions of  both equilibrium and stability problems
as demonstrated below by their comparison with numerics.  Within
the adopted asymptotic approach, the interface elevation
$\eta{\,=\,}\eta_m$ is a new natural variable inherent to the
problem. It first arises explicitly in the equilibrium solution as
the boundary between the extended space charge region and the
region of the diffusion layer, in which the equilibrium states  are
described separately by the outer solution. Then the
 stability study employs perturbation of the interface
elevation $\eta_m$ along with the perturbations of all
conventional variables such as concentrations, velocity etc.
Hence, the system development is accompanied by the interface
instability, which reflects the physical nature of the system.

Following the decomposition method  to leading order in
$\varepsilon$, the new scaled variables $F$ and $\Gamma$  are
introduced for the equilibrium problem:
\begin{equation}\label{eq74c}
F=
\varepsilon\left(\Phi-\frac{J\eta}{2}\right),\ \ \ \Gamma = c^++c^--\frac{1}{2}H^2,\ \ \ \Delta F=
\varepsilon\Delta\Phi,\ \ \ H=\frac{d F}{d\eta}.
\end{equation}

Substituting \eqref{eq74c} into
\eqref{eq41}--\eqref{eq44} yields after some algebra:
\begin{equation}\label{eq71}
\frac{d^2\Gamma}{d\eta^2}=-2\eta\frac{d}{d\eta}
\left(\Gamma+\frac{1}{2}H^2\right),
\end{equation}
\begin{equation}\label{eq72}
\frac{d}{d\eta}\left[\left(\Gamma+\frac{1}{2 }H^2\right)
H\right]=0 ,
\end{equation}
\begin{equation}\label{eq74a}
\left(\Gamma+\frac{1}{2 }H^2\right)H=0,\ \ \ F = 0\ \ \ \mbox{for} \ \ \ \eta=0,
\end{equation}
\begin{equation}\label{eq74b}
\Gamma+\frac{1}{2}H^2\to2,\ \ \ F\to\Delta F\ \ \ \mbox{at}\ \ \ \eta\to\infty,
\end{equation}
and the auxiliary condition for the electric current $J$ related
to $\Gamma$
\begin{equation}\label{JJ}
 J=\frac{d \Gamma}{d\eta},\ \ \ \eta=0.
\end{equation}
To completely determine the outer solution in variables $F$, $H$ and $\Gamma$,
first the patching condition of the outer solutions is applied to
the function $(\Gamma+H^2)H$ at $\eta{\,=\,}\eta_m$, ignoring the input of
the internal boundary layer at $\eta{\,=\,}\eta_m$.
Note also that the method of decomposition ignores the input of the boundary layer in  $\eta=0$ into the outer region, and the boundary condition (29) for $c^+$ at $\eta=0$  is rejected. Then equations \eqref{eq72} and the first of equations \eqref{eq74a} yield
\begin{equation}\label{eq81}
\left(\Gamma+\frac{1}{2}H^2\right)H=0.
\end{equation}
Equation \eqref{eq81} yields the outer solution separately in the space--charge,
$0\leqslant\eta \leqslant \eta_m$, and electro--neutral, $\eta \geqslant \eta_m$, regions:
\begin{equation}\label{REgII}
\Gamma+\frac{1}{2}H^2=0\ \ \ \mbox{for}\ \ \ 0\leqslant\eta < \eta_m,
\end{equation}
\begin{equation}\label{REgIII}
H=0\ \ \ \mbox{for}\ \ \ \eta > \eta_m.
\end{equation}
Then, using equations \eqref{REgII}--\eqref{REgIII}, conditions of
continuity for $F$, $H$, $\Gamma$ and $d\Gamma/d\eta$ and the
rest of equations \eqref{eq71}--\eqref{JJ} lead  to  the following relations in
the space--charge, and electro--neutral regions:
\begin{equation}\label{eq82}
\Gamma=J\left(\eta-\eta_m\right),\ \ \ F=\Delta F\left[1-\left(1-
\frac{\eta}{\eta_m}\right)^{3/2}\right]\ \ \ \mbox{for}\ \ \ 0\leqslant\eta < \eta_m,
\end{equation}
\begin{equation}\label{eq85b}
\Gamma=2-2\frac{\mbox{erfc}(\eta)}{\mbox{erfc}(\eta_m)},\ \ \ F=\Delta F\equiv const\ \ \ \mbox{for}\ \ \ \eta > \eta_m.
\end{equation}
Returning to the old variables, $c^+$, $c^-$, $H$ and $\rho$, we obtain
$$
\displaystyle c^+=c^-=0,\ \ \ H=\sqrt{2J\left(\eta_m-\eta\right)},\ \ \ \rho =\varepsilon \sqrt{\frac{J}{2(\eta_m-\eta)}}\ \ \ \mbox{for}\ \ \ 0\leqslant\eta < \eta_m,
$$
$$
\displaystyle c^+=c^-=1-\frac{ \mbox{erfc}(\eta)}{ \mbox{erfc}(\eta_m)},\ \ \ H=0,\ \ \ \rho\equiv0\ \ \ \mbox{for}\ \ \ \eta > \eta_m.
$$
Here, to the leading order in $\varepsilon$, the volume charge in space--charge region has a singularity at $\eta{\,=\,}\eta_m$. To overcome this inhomogeneity of the outer solution \eqref{eq82}--\eqref{eq85b} in the vicinity of $\eta{\,=\,}\eta_m$, the inner solution should be developed.

According to this solution the potential drop in space--charge region is approximately equal to the total voltage between the membranes:
\begin{equation}
\label{eqPotenDrop}
\Delta F= \varepsilon \Delta\Phi\approx \varepsilon \Delta V.                                                                                     \end{equation}
The values $\Delta F$ and $J$ are interrelated:
\begin{equation}\label{eqDelta F}
\Delta F=\sqrt{\frac{8}{9}J}\,\eta_m^{3/2},
\end{equation}
with the relation between $\eta_m$ and current $J$
\begin{equation}\label{eq87P}
J=\frac{4}{\sqrt{\pi}}\frac{\exp{(-\eta_m^2)}}{\mbox{erfc}\left(\eta_m\right)}.
\end{equation}
Relations  \eqref{eqDelta F} and \eqref{eq87P} determine
implicitly the V--C curve and the  boundary of the ESP region
$\eta_m$ versus potential drop $\Delta F$
\begin{equation}\label{eq88dd}
J\,\mbox{erfc}\left[\left(\frac{9\Delta F^2}{8J}\right)^{1/3}\right]=J_{lim}\exp\left[-\left(\frac{9\Delta F^2}{8
 J}\right)^{2/3}\right],\ \ \ J_{lim}=\frac{4}{\sqrt{\pi}}.
\end{equation}
\begin{equation}\label{eq88d}
\frac{9\sqrt{\pi}}{32}\Delta
F^2=\frac{\eta_m^3\exp\left(-\eta_m^2\right)}{\mbox{erfc}\left(\eta_m\right)},
\end{equation}

\begin{figure}
\begin{center}
    \includegraphics[height=8cm]{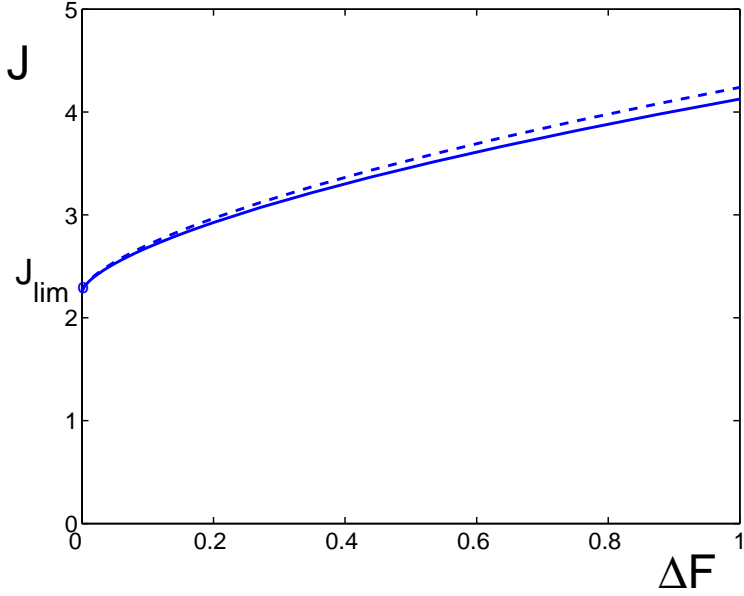}
\end{center}
\caption{VC curves. Solid line relates to \eqref{eq88dd}, while  dashed line -- to the
simplified version \eqref{eq88c}.}\label{VCU}
\end{figure}

In the subsequent stability study of the self--similar regimes, the scaled value of the potential drop is assumed to be small:
\begin{equation}\label{smallDF}
\Delta F\equiv \varepsilon \Delta \Phi \approx \varepsilon \Delta V\ll 1.
\end{equation}
In that limit  relations \eqref{eq88dd}--\eqref{eq88d} are
reduced  as follows:
\begin{equation}\label{eq88e}
\displaystyle J\approx J_{lim}\left(1+\frac{2\eta_m}{\sqrt{\pi}}\right),\ \ \ 
\eta_m \approx \frac{\pi}{8}\left(J-J_{lim}\right),
\end{equation}
\begin{equation}\label{eq88c}
(\Delta F)^2 = \frac{\pi^3}{576}\left(J-J_{lim}\right)^3J.\ \ \ \
\end{equation}
Evidently, (\ref{eq88e})--(\ref{eq88c}) correspond to near--critical currents.
On the other hand the analysis of the inner solution at $\eta=\eta_m$, and of the next order in $\varepsilon$ terms in the outer solution at $\eta>\eta_m$ yields that the approximate solution \eqref{eq82}--\eqref{eq85b} homogeneously satisfies the condition \eqref{eqPotenDrop} of the potential drop domination in  space--charge region if
\begin{equation}\label{DeltaPhi}
\varepsilon\ln \varepsilon^{-1}\ll \Delta F\ll 1.
\end{equation}
Then  relations \eqref{smallDF}--\eqref{DeltaPhi}  gives for a small potential drop:
\begin{equation}\label{eq888dd0}
\varepsilon \ln \varepsilon^{-1} \ll \eta_m^{3/2} \sim(J-J_{lim})^{3/2}\sim\Delta F\ll 1,\,\,\,\,J_{lim}  = O((\Delta F)^0).
\end{equation}
For a small potential drop, relations (\ref{eq82}) and (\ref{eq85b}) yield within space--charge region:
\begin{equation}\label{ESCestim}
c^+= c^-=0,\ \ \ F\sim\Delta F\ \ \ \Gamma\sim(\Delta F)^{2/3},\ \ \ H\sim(\Delta F)^{1/3},\ \ \ \rho\sim(\Delta F)^{-1/3},
\end{equation}
and electro--neutral region:
\begin{equation}\label{EDLestim}
c^+\sim c^-\sim(\Delta F)^0,\ \ \ F\sim\Delta F,\ \ \ \Gamma\sim (\Delta F)^0,\ \ \ H=0,\ \ \ \rho=0.
\end{equation}
while according to equation (\ref{eq888dd0}), the boundary between these two regions is given by
\begin{equation}\label{EDLestimBound}
(\varepsilon \ln \varepsilon^{-1})^{2/3}\ll \eta_m \sim(\Delta F)^{2/3}\ll 1.
\end{equation}

\subsection{Comparison of 1D explicit asymptotic and numeric  solutions }
In  Fig.~\ref{VACM}  the asymptotic solution for V--C curves
 \eqref{eq88dd} is presented as a dependence
$J$  versus  $\Delta \Phi$ (dashed lines)  and compared with those
(solid lines) obtained numerically  in Subsection II.C. The
comparison demonstrates a fair correspondence between the
self--similar asymptotics and exact numerical solution except
the region of under--limiting currents. The difference tends to
zero as $\varepsilon \to 0$ e.g. at $\varepsilon =0.001$ the exact
solution and asymptotics coincide with graphic accuracy. This
justifies the neglect of the EDL inputs into the limiting solutions
accepted in the present study.

\begin{figure}
\begin{center}
    \includegraphics[height=6cm]{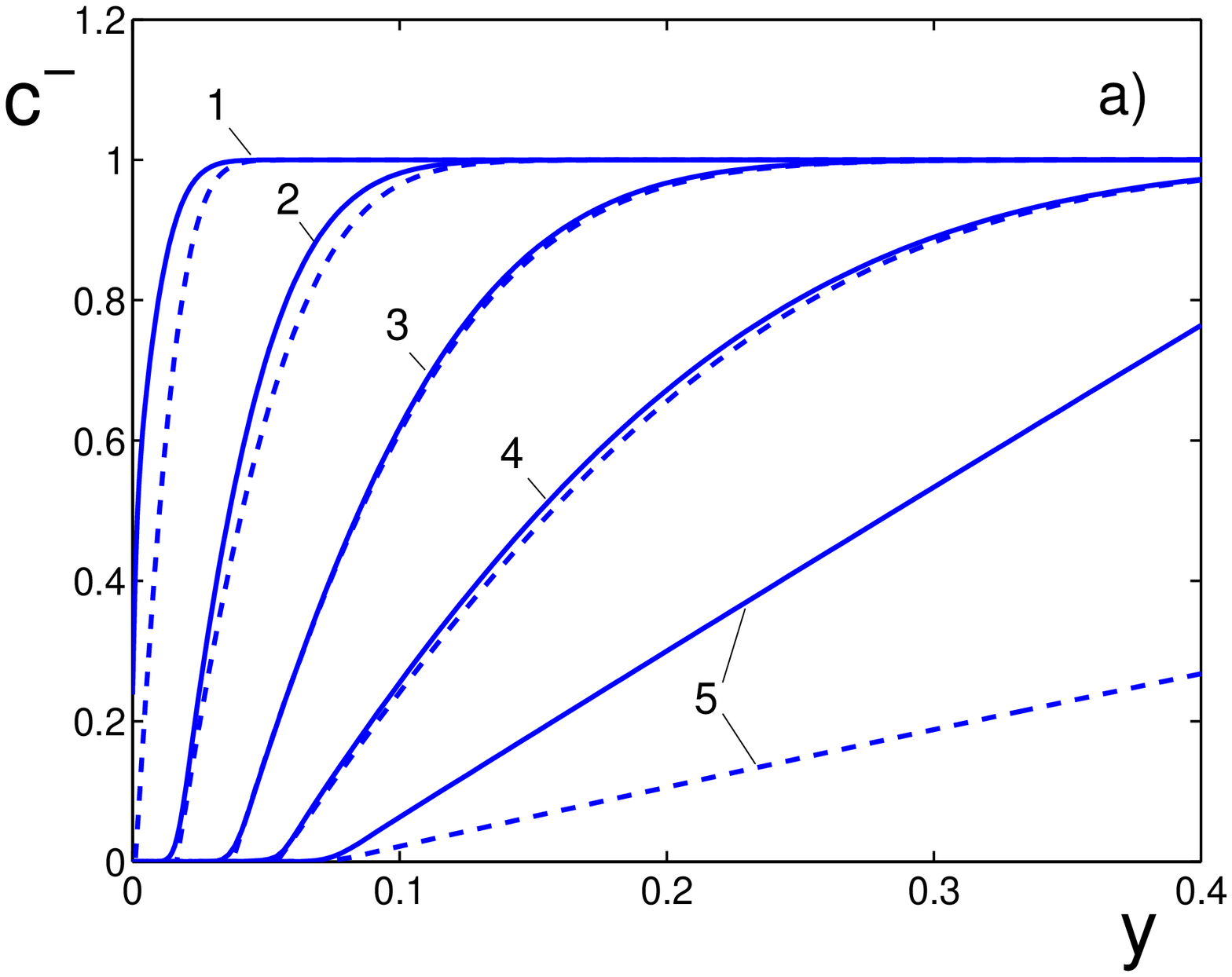}
    \includegraphics[height=6cm]{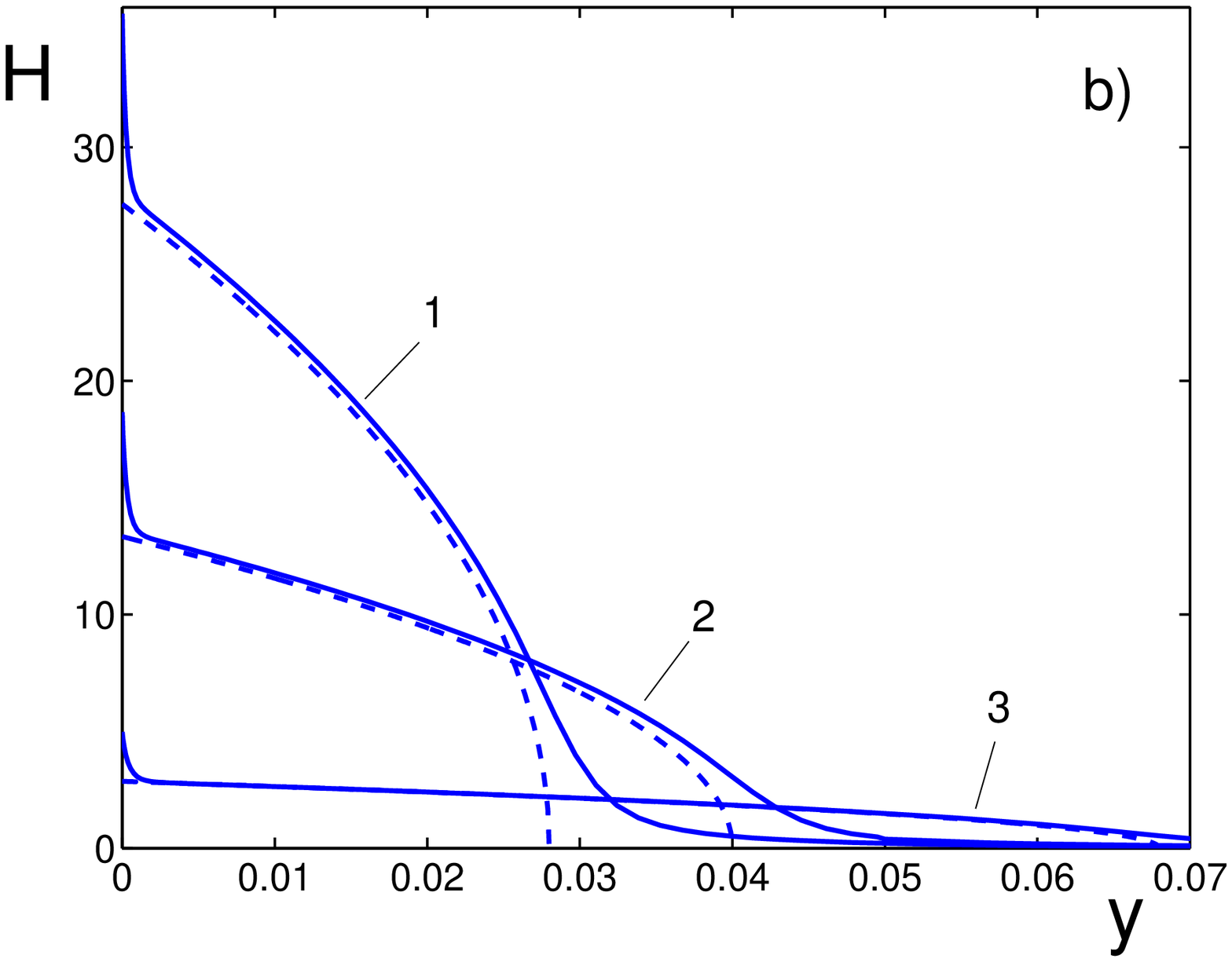}
\end{center}
\caption{Negative ion concentration (a), $c^-$, and electric field (b), $H$,
vs $y$ for several moments of time. Comparison of numerics
\eqref{eqN111}--\eqref{eqJPDim11} (solid lines) with
self--similar asymptotics (\ref{eq82})--(\ref{eq85b})  (dashed
lines) for (a) $\Delta V{\,=\,}50$, $\nu{\,=\,}0.001$: 1.
$t{\,=\,}0.0001$; 2. $t{\,=\,}0.001$; 3. $t{\,=\,}0.004$; 4.
$t{\,=\,}0.015$; 5. $t{\,=\,}0.5$;  and  (b) $\Delta
V{\,=\,}100$, $\nu{\,=\,}0.0005$: 1. $t{\,=\,}0.001$; 2.
$t{\,=\,}0.003$; 3. $t{\,=\,}0.03$.}\label{Cm}
\end{figure}

In Fig.~\ref{VCU} the universal voltage--current characteristic
described by \eqref{eq88dd} is plotted by solid
lines, while simplified version of the VC curve \eqref{eq88c} is
plotted by dashed lines. It is seen that the difference between
these two curves is extremely small for low values of $\Delta F$ and  does not exceed $4\%$ at $\Delta F=1$.

In Fig.~\ref{Cm}  the distributions of the negative ion
concentration $c^{-}$ and electric field $H=\varepsilon E$
obtained from the numerical solution of
\eqref{eqN1111}--\eqref{eq2770P111} are shown for several $t$. In
turn, in Fig.~\ref{Cm}b   self--similar distributions described
by  (\ref{eq82})--(\ref{eq85b})  are given. Except for short and long  times,  a rather good
correspondence is observed of the both approaches at intermediately long
times. In accordance to \eqref{EDLestimBound}, the value of $\eta_m$, indeed, is a small value for parameters adopted in Fig.~\ref{Cm}, where  $\eta_m$ is determined by the condition that $\eta>\eta_m$ correspond to positive values of $c^{-}$.

\begin{figure}
\begin{center}
    \includegraphics[height=8cm]{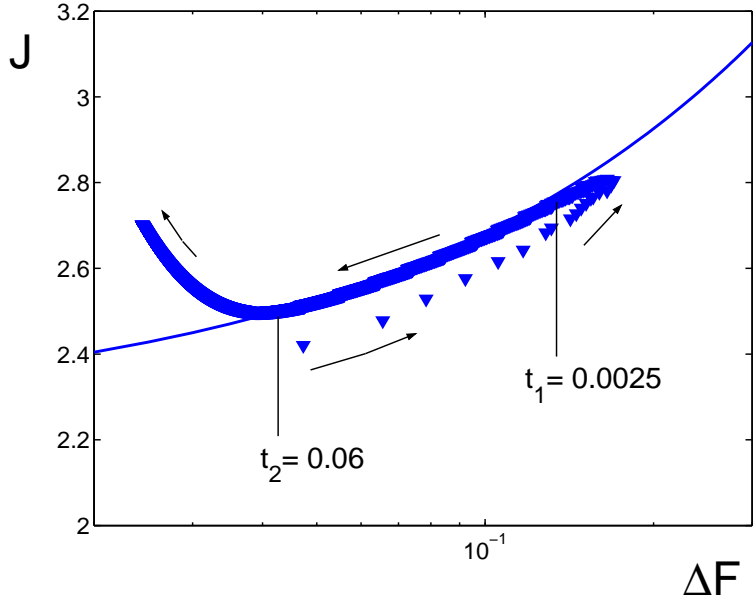}
\end{center}
\caption{Comparison of the V--C characteristics for
self--similar solution and numerics, $\Delta V=50, \nu = 0.00005$.
Numerics \eqref{eqN111}--\eqref{eqJPDim11} is shown by squares and
the universal self--similar VC curve \eqref{eq88dd} -- by solid
line. The region of the solution self--similarity  $t_1{\,<\,t\,<\,}t_2$,
$t_1\gg\nu^2/4 $, $t_2\ll t_s=\tilde{L}^2/(4 \tilde{D})$.
Arrows depict the time growth along the voltage--current curve.}\label{NAJF}
\end{figure}

The self--similar character of the solution for intermediately long times
is illustrated in Fig.~\ref{NAJF}. In this calculation $\Delta V{\,=\,}50,
\nu{\,=\,}0.00005$ are kept constant during  numerical integration of the
system \eqref{eqN1111}--\eqref{eq2770P111}.  Also, the
potential drop between the membrane is kept constant, its fraction
$\Delta \Phi{\,=\,}\Delta V-j(t)/2$ is changing in time as well as
$J$. It is clearly seen that the numerical solution follows the
self--similar equilibrium. This fitting persists up to the moment
$t{\,=\,}t_2$, when the influence of the upper membrane becomes
significant.
\begin{figure}
\begin{center}
    \includegraphics[width=7.5cm]{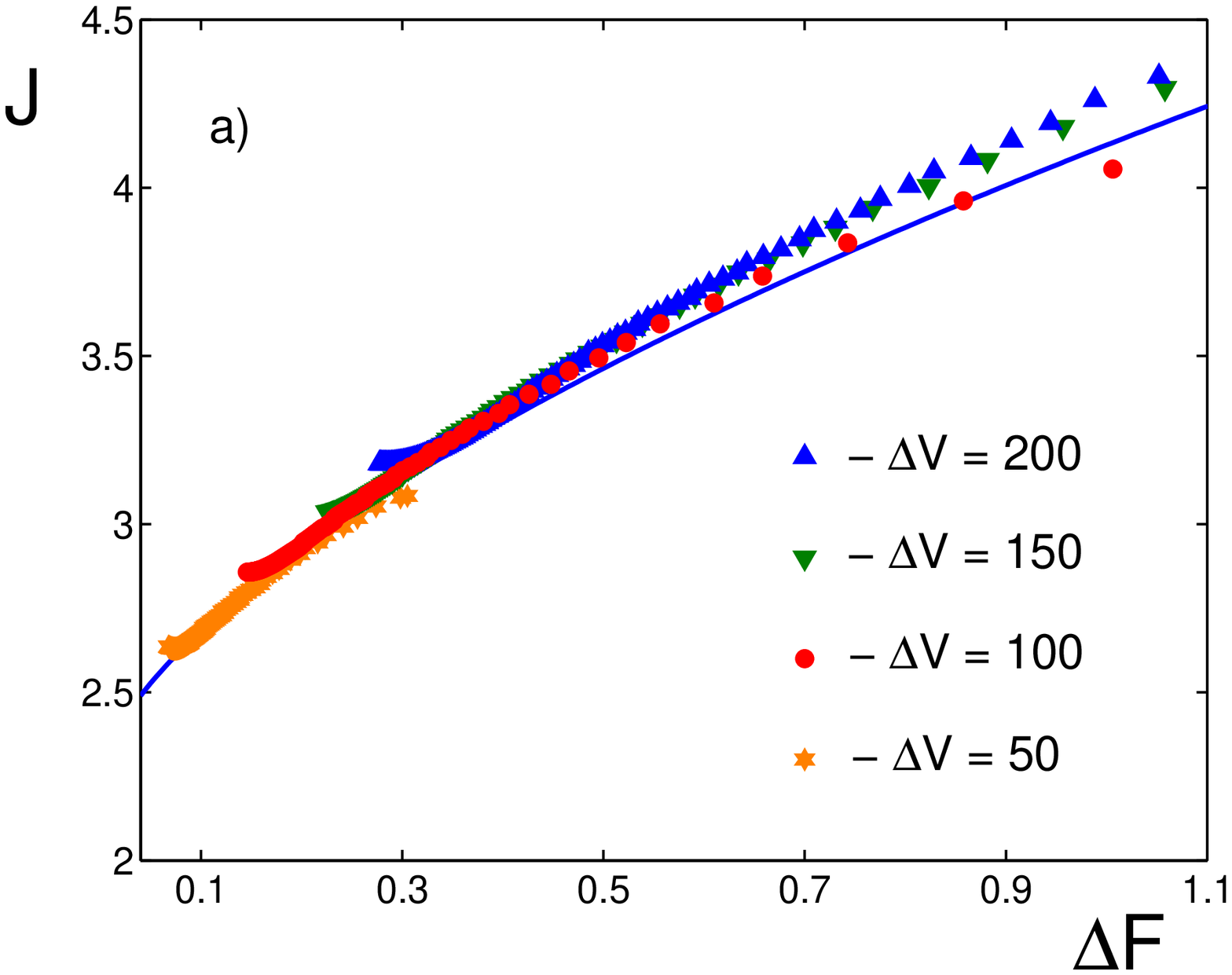}
    \includegraphics[width=7.5cm]{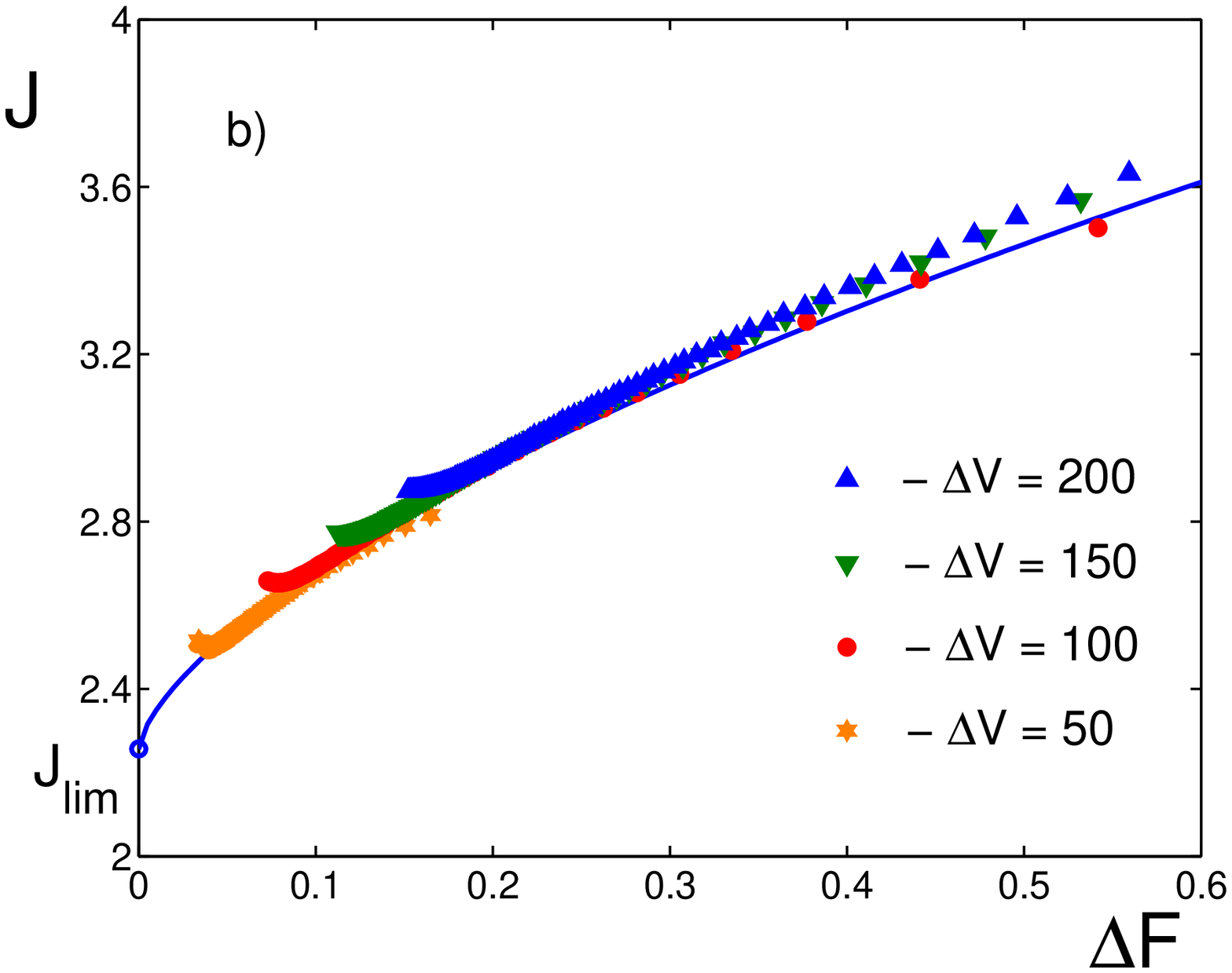}
\end{center}
\caption{Numerical points for several values of the potential drop
between the membranes, $\Delta V$, shrink into the universal  VC curve $J(\Delta F)$ given by
\eqref{eq88dd}; a) $\nu{\,=\,}0.001$, b) $\nu{\,=\,}0.0005$.}\label{CVNUM1}
\end{figure}

The data for several $\Delta V$ are gathered in
Fig.~\ref{CVNUM1}a) and \ref{CVNUM1}b) for $\nu=0.001$ and
$\nu{\,=\,}0.0005$. The data for short times $t{\,<\,}t_1$ and for
long times $t{\,>\,}t_2$ are discarded from the consideration and
only data for intermediately long times are kept. The numerical points
for all $\Delta V$ and $\nu$ shrink into the universal
 VC curve, $J(\Delta F)$, described by \eqref{eq88dd}. It
is seen that the fitting is better for smaller $\nu$ and $\Delta
V$.

\section{2D UNSTEADY PROBLEM FOR SMALL DIMENSIONLESS DEBYE LENGTH }\label{app A}

\subsection{Governing relations.  Dimensionless variables}

Aiming at a  stability study of the 1D
equilibrium self--similar solutions, the 2D  unsteady problem  is rewritten  using self--similar variables.
First, the characteristic  length scales, $\tilde{l}_0$ and $\tilde{\delta}$,  tangential  and normal to the membrane surface
are introduced  together with   the characteristic scales for stream function, velocity components, pressure and electric potential  as follows:
\begin{equation}\label{eq277P3}
 \tilde{l}_0=\frac{1 }{\tilde{\alpha}},\ \ \ \tilde{\delta}=2\sqrt{\tilde{D}\tilde{t}}, \ \ \ 
\tilde{\Psi}_0=\frac{\tilde{D}\tilde{\delta}}{\tilde{l}_0},\ \ \ 
\tilde{U}_0=\frac{\tilde{D}}{\tilde{l}_0},\ \ \
\tilde{V}_0=\frac{\tilde{D}\tilde{\delta}}{\tilde{l}_0^2},\ \ \ 
\tilde{P}_0=\frac{\tilde{\mu}\tilde{D}}{\tilde{\delta}^2},\ \ \ \tilde\Phi_0=\frac{\tilde R\tilde T}{\tilde F}.
\end{equation}
The characteristic time scale, $\tilde{t}_0$,  is also introduced as an arbitrary moderately long   time at which the equilibrium solution is already settled on the self--similar regime and yet does not leave it; $\tilde{\delta}(\tilde{t})$ is the diffusion layer thickness, while $\tilde{\alpha}$ is the characteristic wave
number. Preserving the same notations for dimensionless stream function, velocities and pressure as for their  dimensional values, the following independent self--similar variables are introduced for the 2D
problem:
\begin{equation}\label{SelfSimVar2D}
\tau= \frac{\tilde{t}}{\tilde{t}_0},\ \ \ x=\frac{\tilde{x}}{\tilde{l}_0},\ \ \eta =
\frac{\tilde{y}}{\tilde{\delta}(\tilde{t})}.\ \ \ \
\end{equation}
Then the time--dependent wave number and Debye length are introduced for further convenience
\begin{equation}\label{TimeFuncs}
 \alpha(\tau)= \tilde\delta(\tilde{t})\tilde{\alpha},\ \ \
 \varepsilon(\tau)=\frac{\tilde{\lambda}_D}{
 \tilde{\delta}(\tilde{t})}.
\end{equation}
Functions $\alpha(\tau)$, $\varepsilon(\tau)$, characterizing the time variation of the equilibrium solution, are assumed to be slowly varied in the fast--time scale of perturbations.

Nonlinear 2D problem may be written in the following form:
\begin{equation}\label{InvarEqsKPlus}
\frac{DK}{D\tau}=\nabla\cdot(\rho\nabla \Phi)+\nabla^2K,
\end{equation}
\begin{equation}\label{InvarEqsRhoPlus}
\frac{D\rho}{D\tau}=\nabla\cdot(K\nabla \Phi)+\nabla^2\rho,
\end{equation}
\begin{equation}\label{InvarEqsPhiPlus}
\varepsilon^2\nabla^2\Phi=-\rho,
\end{equation}
\begin{equation}\label{InvarEqsPsiPlus}
\nabla^4\Psi=-\frac{\varkappa}{\varepsilon^2}\nabla'\times(\rho\nabla'\Phi),
\end{equation}
Here
$$
\nabla=\left(\alpha\frac{\partial}{\partial x},\ \frac{\partial}{\partial\eta}\right), \ \ \
\nabla'=\left(\frac{\partial}{\partial x},\ \frac{\partial}{\partial\eta}\right), \ \ \
\frac{D}{D\tau}=4\tau\frac{\partial}{\partial\tau}-2\eta\frac{\partial}{\partial\eta}+\alpha^2\left(U\frac{\partial}{\partial x}+V\frac{\partial}{\partial\eta}\right),
$$
$K = c^++c^-$, $\rho = c^+-c^-$ as previously in 1D problem, $\varkappa=\tilde{\epsilon}\tilde{\Phi}_0^2/(\tilde{\mu}\tilde{D})$
 describes physical properties of the liquid and electrolyte; $\Psi$ is the stream
function:
\begin{equation}\label{eq1012}
U=\frac{\partial\Psi}{\partial\eta},\ \ \ V=-\frac{\partial\Psi}{\partial x}.
\end{equation}

Assuming the problem self--similarity, the following boundary conditions are accepted:
\begin{equation}\label{eq1021a}
\Phi=0,\displaystyle\ \ \ (K-\rho)\frac{\partial\Phi}{\partial\eta}-\frac{\partial
K}{\partial\eta}+\frac{\partial \rho}{\partial\eta}=0,\ \ \ K+\rho=2p,\ \ \ \Psi=\frac{\partial\Psi}{\partial\eta}=0 \ \ \ \mbox{for}\ \ \eta=0,
\end{equation}
\begin{equation}\label{eq1021b}
\displaystyle \Phi-\frac{J\eta}{2}\to\Delta\Phi,\displaystyle\ \ \ \rho\to0,\ \ \ K\to2,\ \ \ \Psi\to0,\ \ \ \frac{\partial\Psi}{\partial\eta}\to0\ \ \  \mbox{at}\ \ \ \eta\to\infty,
\end{equation}
where $J$ is the electric current \eqref{eq44}.

\subsection{Decomposition method for 2D problem. Ad--hoc model}
Extending the decomposition method for the limiting and over--limiting regimes  from 1D case to 2D case,
the following auxiliary variables  scaled in $\varepsilon$ are introduced:
\begin{equation}\label{PhiEFa}
F=\varepsilon\left(\Phi-\frac{J\eta}{2}\right),\ \ \ H=\frac{\partial F}{\partial\eta},\ \ \ \Gamma=K-\frac{1}{2}H^2,\ \ \ \rho=-\varepsilon\left(\frac{\partial H}{\partial\eta}
+\alpha^2\frac{\partial^2 F}{\partial x^2}\right).
\end{equation}

First the system \eqref{InvarEqsKPlus}--\eqref{InvarEqsPsiPlus}  is rewritten in the variables \eqref{PhiEFa} substituting $\rho=-\varepsilon\nabla^2F$:
\begin{equation}\label{InvarEqsK1}
\frac{DK}{D\tau}=-\nabla\cdot[(\nabla^2 F)\nabla F]+\nabla^2K,
\end{equation}
\begin{equation}\label{InvarEqsRho1}
\varepsilon^2\frac{D(\nabla^2 F)}{D\tau}=-\nabla\cdot(K\nabla F)+\varepsilon^2 \nabla^4 F,
\end{equation}
\begin{equation}\label{InvarEqsPsi1}
\nabla^4\Psi=\frac{\varkappa}{\varepsilon^2}\nabla'\times[(\nabla^2 F)\nabla' F],
\end{equation}
and then substituting  $K=\Gamma +\frac{1}{2} H^2$
\begin{equation}\label{InvarEqsK2}
\frac{D(\Gamma +\frac{1}{2}H^2)}{D\tau}=-\nabla\cdot[(\nabla^2 F)\nabla F]+\nabla^2\left(\Gamma +\frac{1}{2} H^2\right),
\end{equation}
\begin{equation}\label{InvarEqsRho2}
\varepsilon^2\frac{D(\nabla^2 F)}{D\tau}=-\nabla\cdot\left[\left(\Gamma +\frac{1}{2} H^2\right)\nabla F\right]+\varepsilon^2 \nabla^4 F.
\end{equation}

The following  ad--hoc model is adopted for the space--charge and electro--neutral regions, respectively:
\begin{equation}\label{ESC_K}
K=\Gamma +\frac{1}{2} H^2\equiv0\ \ \ \mbox{for}\ \ \ 0\leqslant\eta<\eta_m,
\end{equation}
\begin{equation}\label{ESC_H}
F=\Delta F\equiv const\ \ \ (H\equiv0,\ \ \ \rho\equiv 0)\ \ \ \mbox{for}\ \ \ \eta_m>\eta.
\end{equation}
Substituting \eqref{ESC_K} and \eqref{ESC_H} into  \eqref{InvarEqsK1}--\eqref{InvarEqsPsi1} and  \eqref{InvarEqsK2}--\eqref{InvarEqsRho2}, respectively, for the space--charge and electro--neutral regions yield  taking into account the boundary conditions \eqref{eq1021a}--\eqref{eq1021b}:
\begin{equation}\label{ESC_AdHoc}
\Gamma +\frac{1}{2} H^2=0, \ \ \ \nabla\cdot[(\nabla^2F)\nabla F] =0,\ \ \ \nabla^4\Psi=\frac{\varkappa}{\varepsilon^2}\nabla'\times[(\nabla^2F)\nabla' F],\ \ \ \mbox{for}\ \ \ 0\leqslant\eta<\eta_m,
\end{equation}
\begin{equation}\label{EDL_AdHoc}
F=\Delta F\equiv const,\ \ \ \frac{D\Gamma}{D \tau}=\nabla^2 \Gamma,\ \ \ \nabla^4\Psi=0\ \ \ \mbox{for}\ \ \ \eta_m>\eta.
  \end{equation}
The systems \eqref{ESC_AdHoc} and \eqref{EDL_AdHoc} should satisfy the boundary conditions adopted  from \eqref{eq1021a}--\eqref{eq1021b}:
\begin{equation}\label{eqBC_at_0}
F = 0 ,\ \ \ \Psi=\frac{\partial\Psi}{\partial\eta}=0 \ \ \ \mbox{for}\ \ \ \eta=0,
\end{equation}
\begin{equation}\label{eqBC_at_infy}
\Gamma\to2,\ \ \ \Psi\to0,\ \ \ \frac{\partial\Psi}{\partial\eta}\to0\ \ \ \mbox{at}\ \ \ \eta\to\infty.
\end{equation}
As in 1D model, the method of decomposition ignores the input of the boundary layer in the near--bottom region into the outer region, and rejects  the boundary conditions  for $c^+=(K+\rho)/2$ at $\eta=0$ in \eqref{eq1021a}.
Additionally, the problems   \eqref{ESC_AdHoc} and  \eqref{EDL_AdHoc}  should be complemented by the proper patching conditions which provide continuity of the solutions  of the systems \eqref{ESC_AdHoc} and  \eqref{EDL_AdHoc} at the patching point  $\eta=\eta_m$:
\begin{equation}\label{eqBC_at_eta_m}
\ \ \Gamma_-= \Gamma_+=0,\ \ F_-=F_+= \Delta F, \ \
\nabla_n F_-= 0,\ \  \nabla_n \Gamma_-= \nabla_n \Gamma_+,
 \  \
 \nabla^k_n  \Psi_-= \nabla^k_n  \Psi_+,
\end{equation}
where $\nabla_n$ is the gradient projection to the  direction normal to the interface $\eta=\eta_m$;
 $\Delta F\equiv const$; $k=0,1,2,3$  due to the fourth order in $\eta$ of the differential equations
 for $\Psi$; subscripts $-$ and $+$ denote the values of the corresponding variables on the
 left and right  of $\eta=\eta_m$, respectively (for details see ~\cite{Disser}).

The resulting non--linear model consists of a small parameter $\varepsilon$, but the linearized problem is further simplified by an additional rescaling in the electroneutral region that excludes  $\varepsilon$ from the linear problem. In the next section the stability problem obtained by linearization about the self--similar 1D solution is considered. As is clear from the above analysis of the 1D self--similar problem, the resulting unsteady problem  depends on two small  dimensionless parameters: Debye length and total potential drop, and  further  modeling is restricted by small values of the potential drop. Finally, the results for the asymptotic model are compared with direct numerical simulations for the exact problem.

\section{STABILITY  OF SELF--SIMILAR SOLUTIONS}
The developed 1D solution can lose stability, and a new
electro--convective regime bifurcates at some values of
parameters. Let us restrict ourselves with the marginal stability
problem,
$$
\frac{\partial}{\partial\tau}=0.
$$
and impose infinitesimal perturbations on the self--similar 1D
solution. Using the fact that the coefficients of equations
\eqref{ESC_AdHoc}--\eqref{EDL_AdHoc} are slowly changing at long
times:
\begin{equation}\label{DistPsi1}
R = \bar{R}(\eta) + \hat{R}(\eta)\exp{(ix)},
\end{equation}
where $R$ stands for any physical variables, e.g. for $\Gamma$,
$H$, $F$, $\eta_m$, $J$ or $\Psi$. Bars and hats denote the unperturbed and perturbed variables, respectively (below bars for the unperturbed variables are dropped wherever possible with no confusion).

\subsection{Space charge region, $0 \leqslant \eta < \eta_m$.}
 {\it{Electrostatic problem.}}
 Substituting the relations \eqref{eq82}
  and \eqref{DistPsi1} into
\eqref{ESC_AdHoc} and linearizing with respect to
small perturbations $\hat{\Gamma},\ \hat{H},\ \hat{F},\
\hat{\eta}_m,\ \hat{J}$, yields the following  system  in the
space charge region, $0 \leqslant \eta < \eta_m$:
\begin{equation}\label{eq1118}
\displaystyle
\hat{\Gamma}+H\frac{d \hat{F}}{d \eta}=0,
\end{equation}
\begin{equation}\label{eq1119}
\displaystyle \frac{d^2\hat{\Gamma}}{d
\eta^2}+\alpha^2\left(H\frac{d \hat{F}}{d \eta}+2\frac{d
H}{d \eta}\hat{F}\right)=0.
\end{equation}
 Linearizing  the BC's \eqref{eqBC_at_0},
\eqref{eqBC_at_eta_m}  and shifting them to the undisturbed boundary
$\eta=\eta_m$, yields the boundary conditions for $\hat{F}$
and $\hat{\Gamma}$:
\begin{equation}\label{eq1120}
\begin{array}{c}
\displaystyle\hat{F}=0,\ \ \frac{d\hat{\Gamma}}{d\eta}=\hat{J} \ \ \ \mbox{for}\ \ \ \eta=0, \\[12pt]
\displaystyle\displaystyle\hat{F}=0,\ \ \ \hat{\Gamma}+J\hat{\eta}_m=0\ \ \ \mbox{for} \ \ \ \eta=\eta_m.
\end{array}
\end{equation}
Four conditions  \eqref{eq1120} serve the boundary conditions for
the third order system \eqref{eq1118}--\eqref{eq1119} and
determine the disturbed electric current $\hat{J}$.
After eliminating $\hat{\Gamma}$, our system
\eqref{eq1118}--\eqref{eq1120} turns into the boundary problem for
$\hat{F}$:
\begin{equation}\label{eq1123}
\displaystyle
\frac{d^2}{d\eta^2}\left(H\frac{d\hat{F}}{d\eta}\right)-\alpha^2
\left(H\frac{d\hat F}{d\eta}+2\frac{dH}{d\eta}\hat F\right)=0,
\end{equation}
\begin{equation}\label{eq1124}
\begin{array}{c}
\hat{F}=0,\ \ \ \displaystyle
\frac{d}{d\eta}\left(H\frac{d\hat{F}}{d\eta}\right)=-\hat{J}\ \ \ \mbox{for}\ \ \ \eta=0,\\[10pt]
\hat{F}=0,\ \ \ \displaystyle
H\frac{d\hat F}{d\eta}=J\hat\eta_m\ \ \ \mbox{for}\ \ \ \eta=\eta_m.
\end{array}
\end{equation}
Our knowledge of the solution is complemented by the following
expression for the charge density perturbation, $\hat{\rho}$:
\begin{equation}\label{distRho}
\hat{\rho} =
-\varepsilon\left(\frac{d^2\hat{F}}{d\eta^2}-\alpha^2\hat{F}\right).
\end{equation}

{\it{Hydrodynamic  problem.}}
The linearized  hydrodynamic problem in the region $0\leqslant\eta<\eta_m$, subjected to the boundary conditions \eqref{eqBC_at_eta_m} at $\eta=\eta_m$
and at $\eta=0$, is as follows:
\begin{equation}\label{eq1127}
\frac{d^4\hat{\Psi}}{d\eta^4}-2\alpha^2\frac{d^2\hat{\Psi}}{d\eta^2}+\alpha^4\hat{\Psi}
=\frac{i\varkappa}{\varepsilon^2}\left(H
\frac{d^2\hat{F}}{d\eta^2}-\frac{d^2H}{d\eta^2}\hat{F}-\alpha^2H\hat{F}\right),
\end{equation}
\begin{equation}\label{BC_0}
 \hat{\Psi}=\frac{d\hat{\Psi}}{d\eta}=0 \ \ \ \mbox{for}\ \ \ \eta=0,
 \end{equation}
\begin{equation}\label{BC_eta_m}
\frac{d^2\hat{\Psi}}{d\eta^2}+2\alpha\frac{d\hat{\Psi}}{d\eta}+\alpha^2\hat{\Psi}= 0,\ \ \ \frac{d^3\hat{\Psi}}{d\eta^3}-3\alpha^2\frac{d\hat{\Psi}}{d\eta}-2\alpha^3\hat{\Psi}= 0\ \ \ \mbox{for}\ \ \ \eta=\eta_m.
\end{equation}

Solution of the problem  \eqref{eq1127}--\eqref{BC_eta_m}  determines the perturbations of the slip velocity
\begin{equation}\label{eqUmVm}
\hat{U}_m=\frac{d\hat{\Psi}}{d\eta},\ \ \ \hat{V}_m = -i\hat{\Psi}\ \ \ \mbox{for}\ \ \ \eta=\eta_m.
\end{equation}

Solutions of  the electrostatic and hydrodynamic problems in the space charge region are presented in
Appendixes A and B. In particular, the scaled slip velocities, $u$ and $v$ are depicted in  Fig.~\ref{uv}
\begin{equation}\label{scaled slip}
\hat{U}_m=-i\alpha \hat Cu\left(r\right), \ \ \  \hat{V}_m=\hat Cv\left(r\right).
\end{equation}
where  $\hat C$ is a constant factor (see Appendix B). For small $r$
tangential component of velocity coincides with obtained in
\cite{Rub3a}. The values of the components of the scaled slip
velocities at $\eta=\eta_m$ will be used as boundary conditions
for the hydrodynamic problem in the electro--neutral region  $\eta
> \eta_m$.

\begin{figure}
  \begin{center}
    \includegraphics[height=8cm]{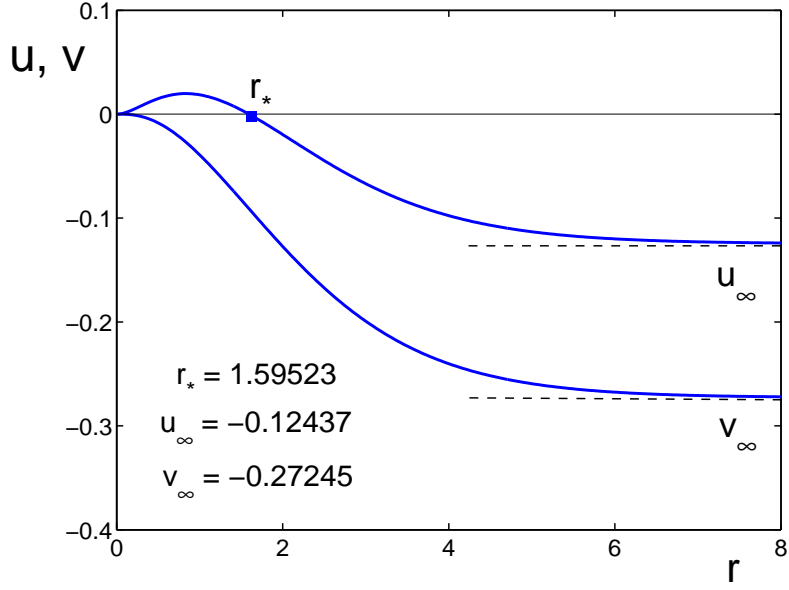}
  \end{center}
\caption[]{Components of scaled slip velocities $u$ and $v$.}\label{uv}
\end{figure}

\subsection{Electro--neutral region, $\eta > \eta_m$.}

Linearizing the problem \eqref{EDL_AdHoc}, \eqref{eqBC_at_infy}, \eqref{eqBC_at_eta_m}  about the
1D self--similar solution \eqref{eq85b} yields the eigenvalue
stability problem in the electro--neutral region $\eta>\eta_m$.

 {\it{Electrostatic  problem.}}
\begin{equation}\label{eq1509}
\displaystyle
\frac{d^2\hat{\Gamma}}{d\eta^2}+2\eta\frac{d\hat{\Gamma}}{d\eta}-\alpha^2\hat{\Gamma}
=-i\alpha^2J\hat{\Psi}\left(\eta\right)e^{\eta_m^2-\eta^2},
\end{equation}
\begin{equation}\label{eq1510}
\displaystyle
\frac{d^4\hat{\Psi}}{d\eta^4}-2\alpha^2\frac{d^2\hat{\Psi}}{d\eta^2}+\alpha^4\hat{\Psi} = 0,
\end{equation}
The problem is completed by the following boundary conditions:
\begin{equation}\label{eq1511}
\displaystyle\hat{\Gamma}+J\hat{\eta}_m=0,\ \ \ \frac{d\hat{\Gamma}}{d\eta}=\left.\frac{d\hat{\Gamma}}{d\eta}\right|_{\eta=\eta_m-0}+2J\eta_m\hat{\eta}_m,\ \ \ 
\hat{\Psi}=i\hat{V}_m,\ \ \ \frac{d\hat{\Psi}}{d\eta}=\hat{U}_m\ \ \ \mbox{for}\ \ \ \eta=\eta_m,\\
\end{equation}
\begin{equation}\label{eq1511Plus}
\displaystyle \hat{\Gamma}\to0,\ \ \ \hat{\Psi}\to0\ \ \ \mbox{at}\ \ \ \eta\to\infty,
\end{equation}
where the slip velocity $\hat{U}_m$, $\hat{V}_m$ is taken from the solution of the problem in the
region  $0 \leqslant  \eta < \eta_m$ at
$\eta=\eta_m$.

{\it{Hydrodynamic problem.}} Notice that the hydrodynamical part
of the problem can be solved separately and independently from the
rest part of the problem. Solution in the space--charge region is
presented in Appendix B, Eqs.~\eqref{psia},
\eqref{psiseries}\textbf{--}\eqref{AB}
 (for details see also~\cite{Disser}).
The velocity distribution at $\eta_m{\,<\,\eta\,<\,}\infty$ can be
found by solving \eqref{eq1510} with the boundary conditions
\eqref{eq1511} and \eqref{eq1511Plus}. Solution is found using
\eqref{eqUmVm} and \eqref{scaled slip},
\begin{equation}\label{psis}
\begin{array}{c}
\hat{\Psi}=i\hat C\psi,\\[10pt]
\psi=
\left\{
\begin{array}{ll}
\displaystyle \left[v+(u-v)\xi\right]e^\xi+\psi_p(\xi)\,,&\ \ \ 0<\eta\leqslant\eta_m,\\[12pt]
\displaystyle \left[v+(u-v)\xi\right]e^\xi,&\ \ \ \eta>\eta_m,
\end{array}
\right.
\end{array}
\end{equation}
where $\xi{\,=\alpha(\eta_m-}\eta)$, the values $\hat C$, $u$,
$v$, $\psi_p$ are calculating according to Eqs.~\eqref{psia},
\eqref{u}\textbf{--}\eqref{eq1418}.

Eventually, the stream--lines determined as the level curves of the imaginary part of the
perturbed stream function  can be built up
\begin{equation}\label{psisa}
\Psi = \hat{\Psi}(\eta)\exp(ix),\ \ \ \hat{\Psi} = i\hat C\psi,
\end{equation}
\begin{equation}\label{psisb}
\mathcal{R}e\left(\Psi\right) = -2\hat C\psi(\eta)\sin(x) = c(\tau).
\end{equation}
Note that the stream--function lags behind the charge density
by $\pi/2$, while the charge density $\hat{\rho}$ is in a counter phase with the interface elevation $\hat{\eta}_m$. Since the function $\hat{\Psi}$ has a single extremum
point and  vanishes at $0$ and  infinity, the stream--lines have
closed trajectories corresponding to 2D electro--convective rolls.
The calculation results for
 $\Delta\Phi{\,=\,}100,\ \varepsilon{\,=\,}0.001$
are presented in Fig.~\ref{Stream} a)--f). For increasing $\alpha$,
the centers of the rolls arriving from the infinity are crossing
the boundary of the space charge region, and then, at
$\alpha\rightarrow\infty$, are approaching the boundary
$\eta=\eta_m$ from below.

\begin{figure}
  \begin{center}
    \includegraphics[ width=7.4cm]{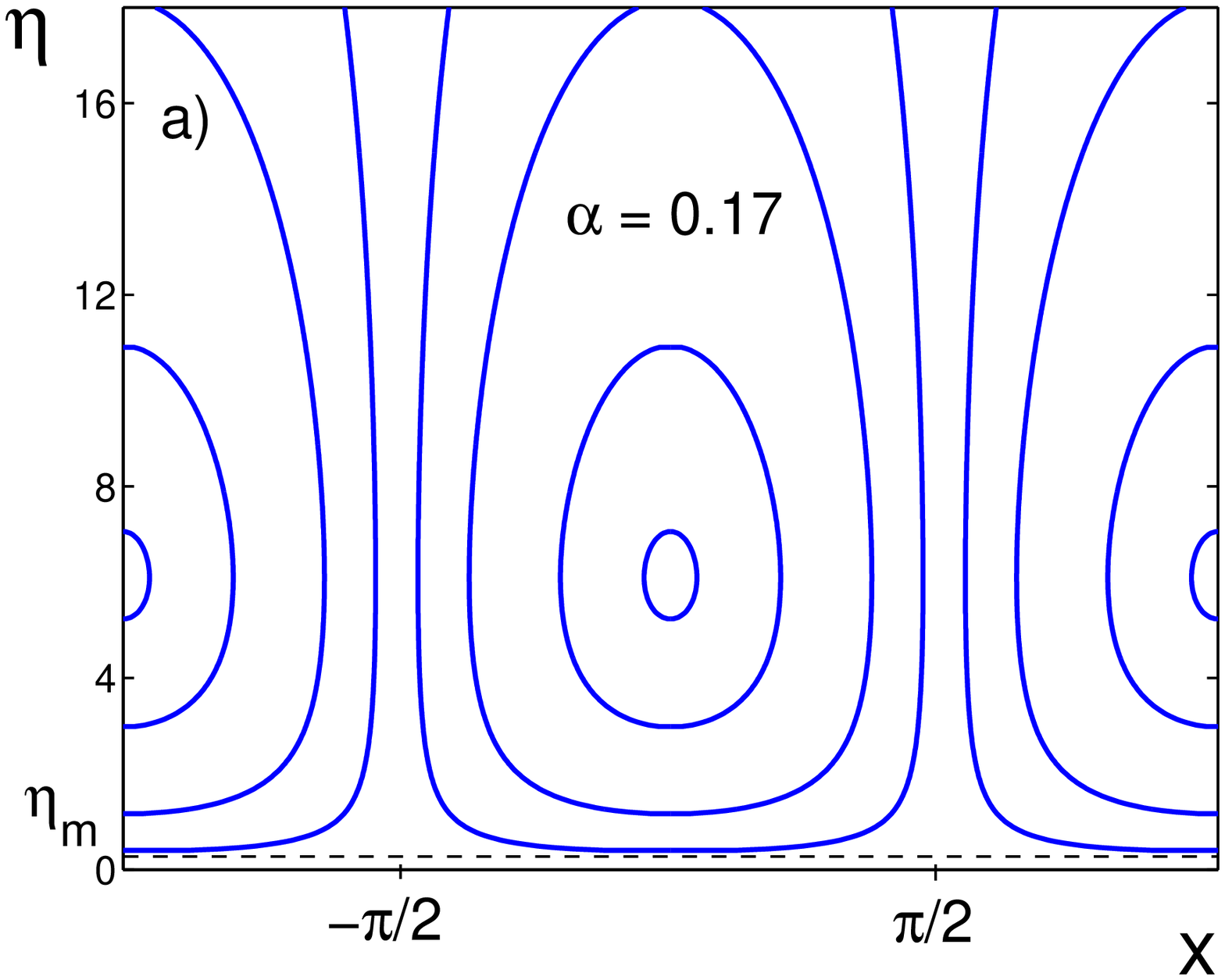}
    \includegraphics[width=7.4cm]{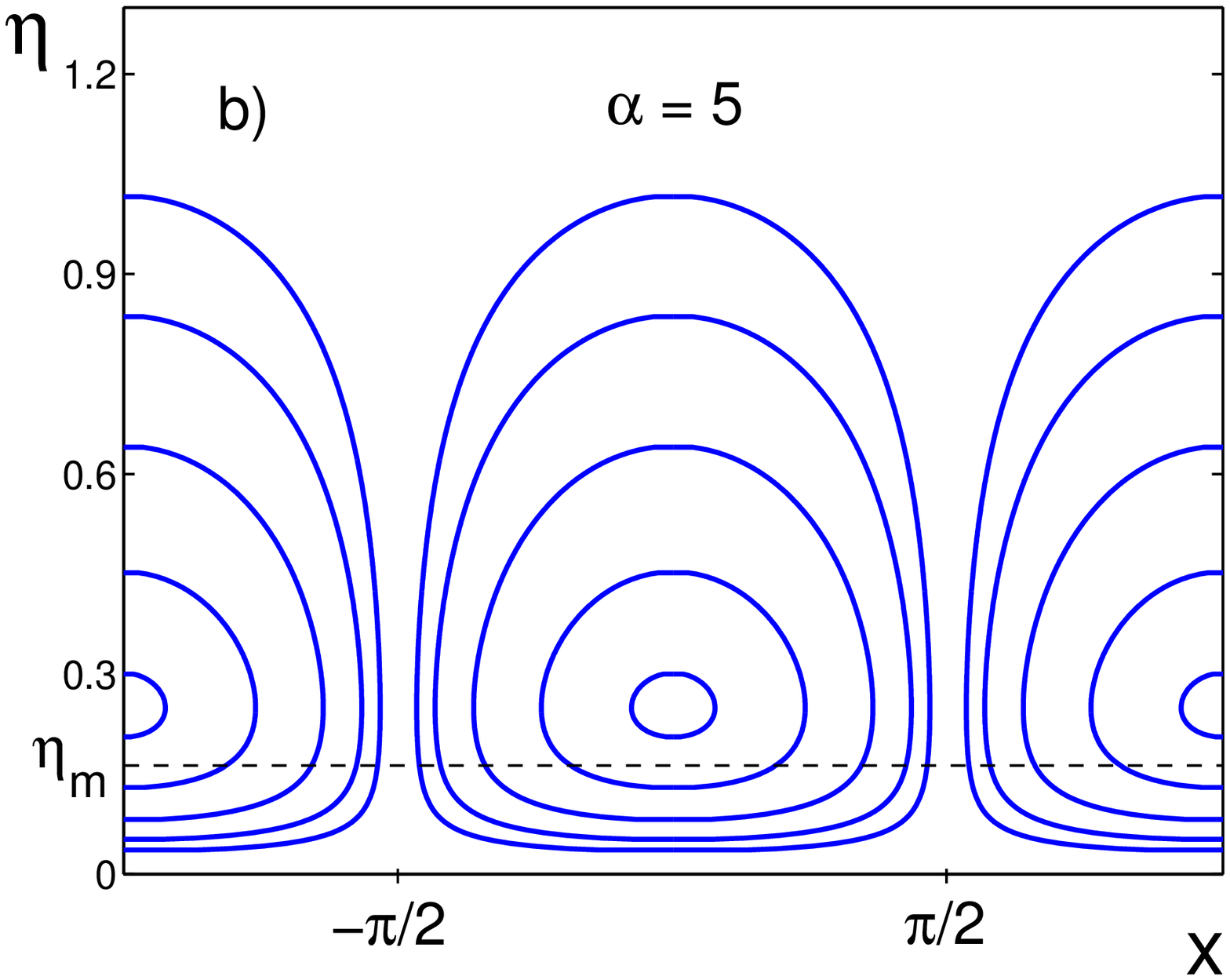}
  \end{center}
    \begin{center}
    \includegraphics[width=7.4cm]{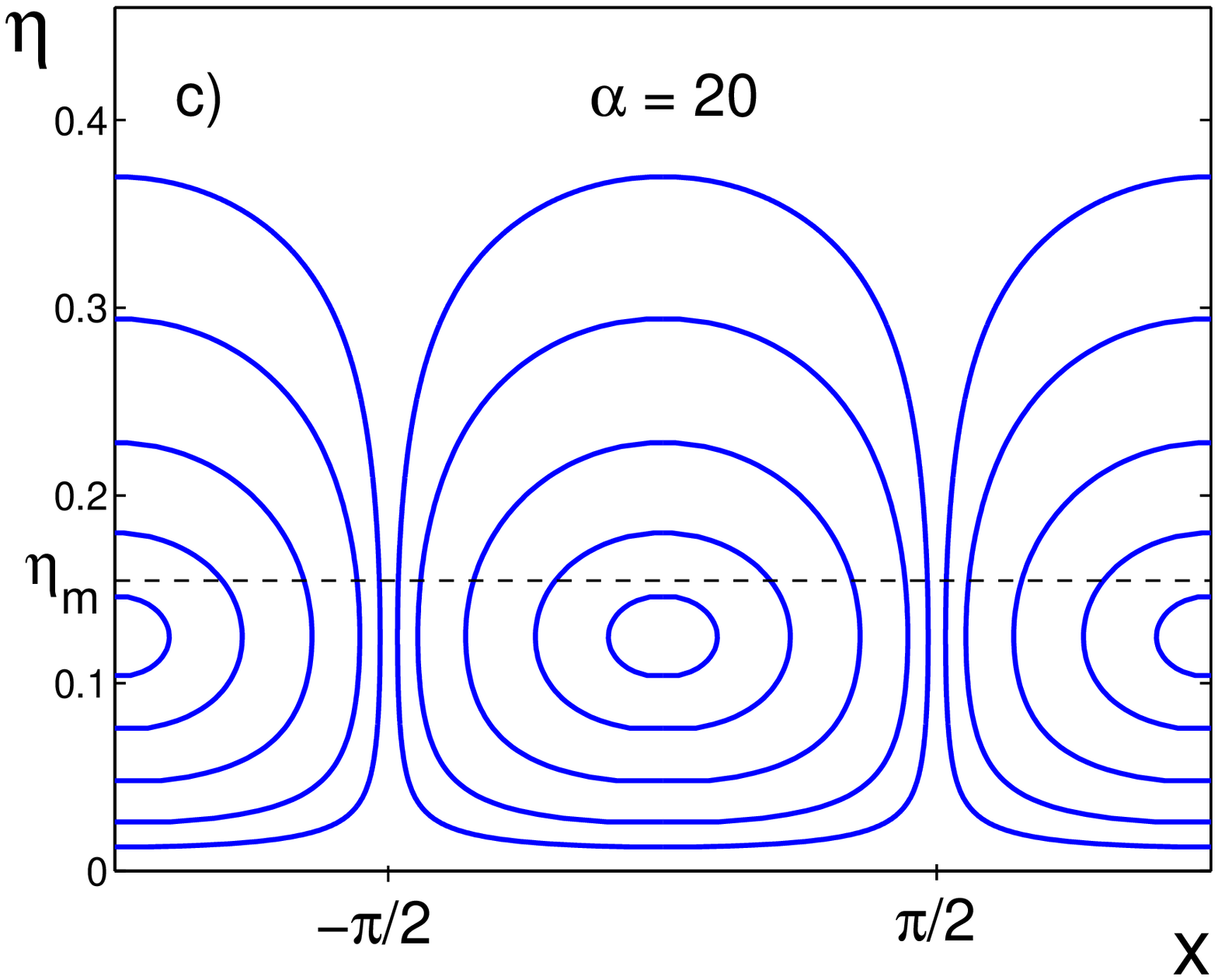}
    \includegraphics[width=7.4cm]{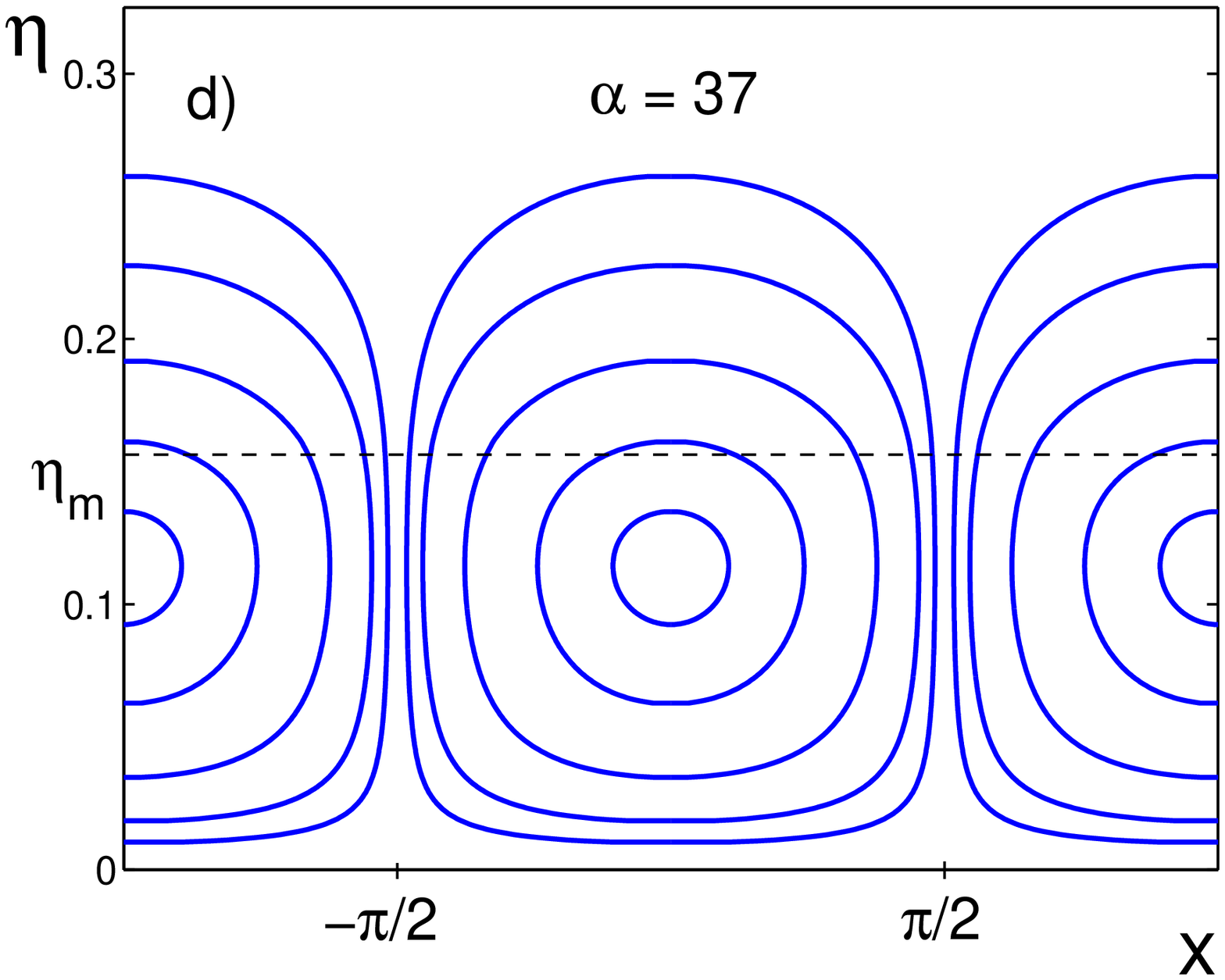}
  \end{center}
    \begin{center}
    \includegraphics[width=7.4cm]{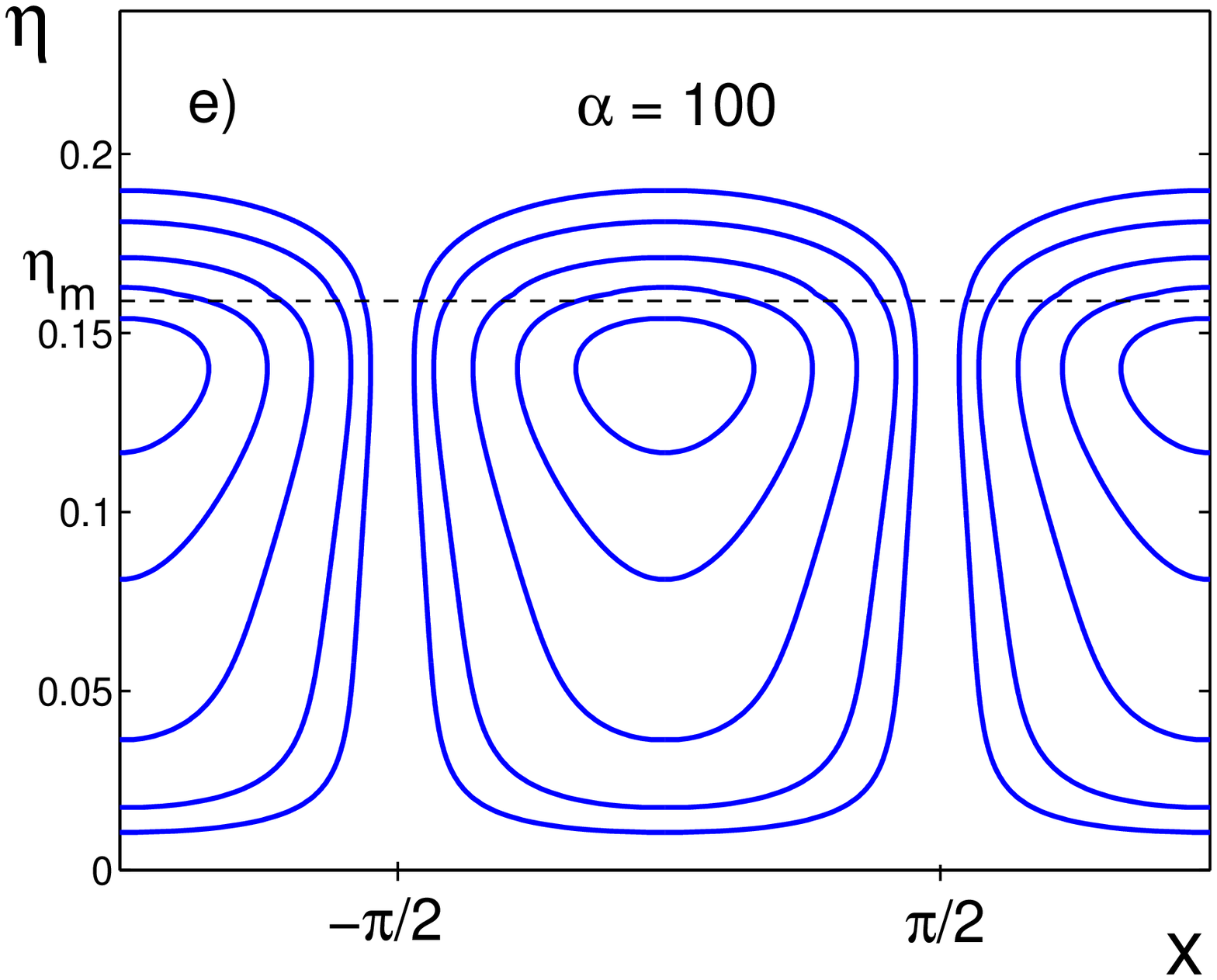}
    \includegraphics[width=7.4cm]{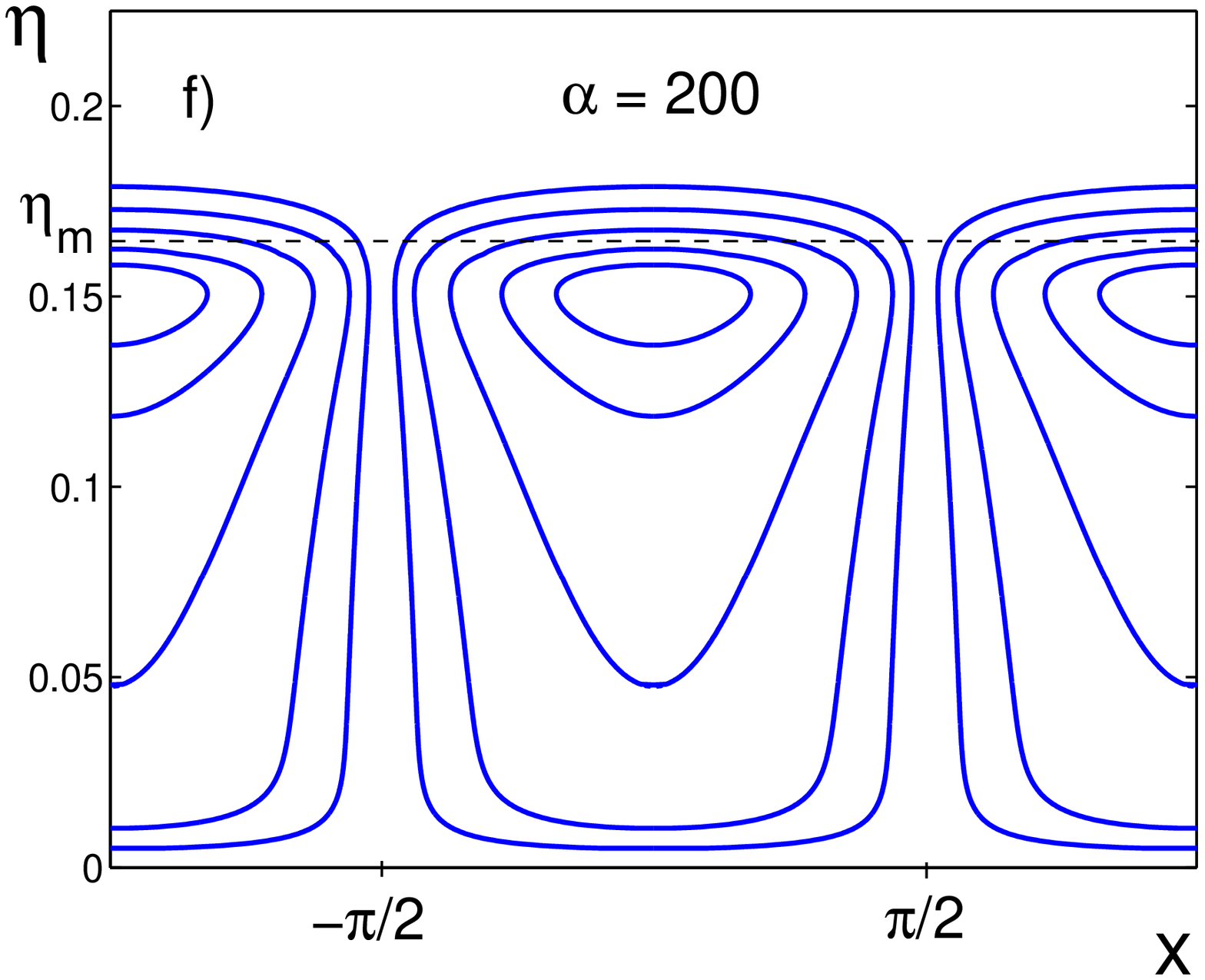}
  \end{center}
\caption[]{Stream--lines for $\Delta\Phi{\,=\,}100,\ \varepsilon{\,=\,}0.001$
and different values of $\alpha$.}\label{Stream}
\end{figure}

\subsection{Results for the  linear stability analysis.}
The final step of our analysis is to find $\hat{\Gamma}$ in the
electro--neutral region, $\eta>\eta_m$, and solve the stability
problem. Boundary conditions for $\Gamma$ at $\eta=\eta_m$ are
 $$
\hat\Gamma=\frac{G(r)}{\alpha}\hat J,\ \ \ \frac{d\hat{\Gamma}}{d\eta}=-\left(3\alpha a\left(r\right)+2\eta_m\right)\frac{G(r)}{\alpha}\hat{J}\ \ \ \mbox{for}\ \ \ \eta=\eta_m.
$$
 Further it will be suitable to normalize 
$\hat\Gamma$ to $G\hat{J}/\alpha$,
\begin{equation}\label{eq1515}
\displaystyle \frac{d^2 \hat{\Gamma}}{d \eta^2}+2\eta\frac{d\hat{\Gamma}}{d\eta}-\alpha^2\hat{\Gamma}=
\frac{2\varkappa J}{\alpha\varepsilon^2}\:e^{\alpha\eta_m+\eta_m^2}\left(v+\alpha\left(u-v\right)(\eta_m-\eta)\right)e^{-\alpha\eta-\eta^2},
\end{equation}
\begin{equation}\label{eq1516}
\displaystyle \ \ \displaystyle\hat{\Gamma}=1,\ \ \ \frac{d\hat{\Gamma}}{d\eta}=-3\alpha a-2\eta_m\ \ \ \mbox{for}\ \ \ \eta=\eta_m\\[12pt],
\end{equation}
\begin{equation}\label{eq1516Plus}
\hat{\Gamma}\to0 \ \ \ \mbox{at}\ \ \ \eta\to\infty.
\end{equation}

The present asymptotic analysis  in small  $\varepsilon $ for
the limiting regimes has no restrictions for the value of
$\alpha$. The system \eqref{eq1515}--\eqref{eq1516} is solved
numerically by the shooting method, and the calculated marginal
stability curves are presented in Fig.~\ref{NSTnum}. In this
picture results of the numerical analysis without
any restriction on $\alpha$ and $\varepsilon$ are presented \cite{Dem}. The
error of the present asymptotics can be significant near the nose
of the marginal stability curves, but  small far from it. This
occurs since the nose of the marginal curve is located near the
transition region from the under--limiting to limiting regime (see
Fig.~\ref{PAJ}), where our asymptotics is not accurate.

 Since the dimensionless Debye thickness and wave number
 are slow functions of time at moderately long times, we obtain, using for clearness the dimensional variables (denoted by tilde):
$$
\varepsilon=\frac{\tilde{\lambda}_D}{\sqrt{4 \tilde{D} \tilde{t}}}, \
\ \ \ \alpha=\tilde{\alpha} \sqrt{4 \tilde{D} \tilde{t}}.
$$
Note that the coefficients of the eigenvalue problem
\eqref{eq1515}--\eqref{eq1516} are slowly varied with time. Note
also that in spite of their slow changing, the parametric
dependence of $\alpha$ and $\varepsilon$ on time significantly
changes the interpretation of the marginal stability curves.
Taking this into account and following Shtemler (1981) \cite{Sht}, we present
our neutral curves in coordinates $1/\varepsilon$ and $\alpha$,
taking $\Delta\Phi$ as a constant parameter. The instantaneous
values of $1/\varepsilon$ and $\alpha$ depend on time implicitly,
$\alpha \sim \sqrt{4 \tilde{D} \tilde{t}}$ and $1/\varepsilon \sim
\sqrt{4 \tilde{D} \tilde{t}}$ with the constant ratio
$$
k=\frac{\alpha}{1/\varepsilon}=\frac{\tilde{\alpha}}{1/\tilde{\varepsilon}}
= \tilde{\alpha}\tilde{\lambda}_D
$$
 Hence, any straight line which starts in
the origin of the plane  $(1/\varepsilon,\ \alpha)$ with inclination
$k$
\begin{equation}\label{eq15211}
\alpha=k \frac{1}{\varepsilon}
\end{equation}
 characterizes wave number $\tilde{\alpha}$ constant with respect to time. If time
 is increasing, the instantaneous values of $1/\varepsilon$ and $\alpha$
 lying on this line are increasing as $\sqrt{4 \tilde{D} \tilde{t}}$.
 For large enough $k$, the whole straight line is in the
 stable region. Then, if we decrease the tangent $k$, at some $k=k_0$
 the line \eqref{eq15211} will be tangential to the neutral
curve (see Fig.~\ref{Alpeps},  upper left corner). The dependences of
$k_0$, $\alpha_0=\alpha(k_0)$ and $1/\varepsilon_0=1/\varepsilon(k_0)$
as functions of $\Delta \Phi$ are given in Table \ref{T1}.

\begin{figure}
  \begin{center}
    \includegraphics[height=8cm]{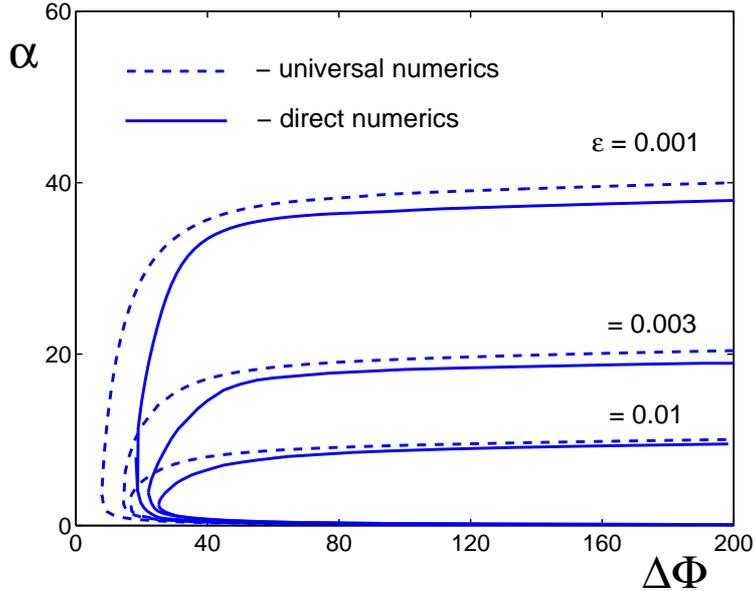}
  \end{center}
\caption[]{Marginal stability curves for self--similar
asymptotic in $\varepsilon$ solution \eqref{eq1515}--\eqref{eq1516} (dashed lines) and exact numerical
solution (solid lines) \cite{Dem}.}\label{NSTnum}
\end{figure}
\begin{figure}
  \begin{center}
    \includegraphics[height=8cm]{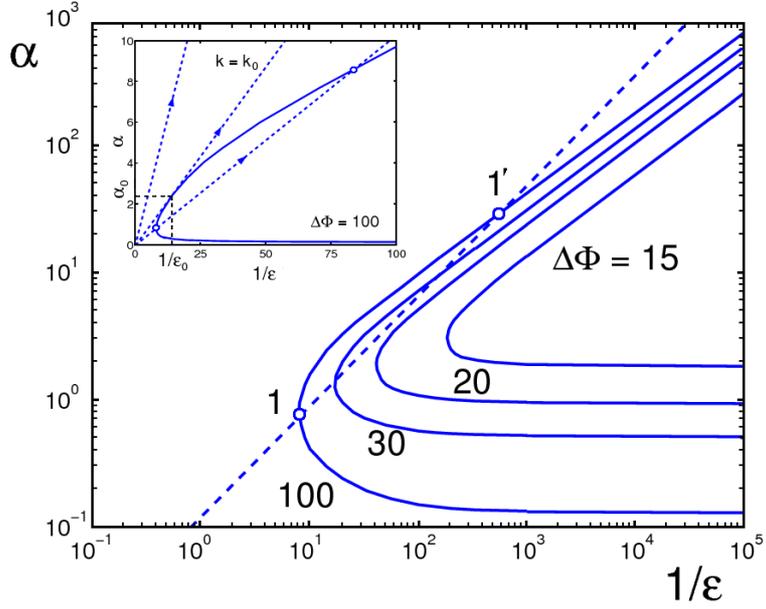}
  \end{center}
\caption[]{Marginal stability curves in $1/\varepsilon$ and $\alpha$
coordinates.} \label{Alpeps}
\end{figure}

With further decrease of  $k < k_0$, the straight line crosses
the neutral curve in two points, $1$ and $1'$, and the region along
the line can be divided into three regions:  I -- the region
between the origin and point $1$; II -- the region within the
neutral curve, between points $1$ and $1'$; III -- the region lying
above the upper branch of the neutral curve.
\begin{table}
\begin{center}
\begin{tabular}{|p{1.5cm}|p{1.5cm}|p{1.5cm}|p{1.5cm}|p{1.5cm}|p{1.5cm}|}
\hline \centering $\Delta\Phi$ & \centering $k_0$ & \centering $\alpha_0$ & \centering $1/\varepsilon_0$ & \centering $Q(\Delta\Phi)$ & \ $R(\Delta\Phi)$ \\
\hline \centering $15$ & \centering $0.02$ & \centering $4.38$ & \centering $218.8$ & \centering $0.14$ & \ \ $1.83$ \\
\hline \centering $20$ & \centering $0.06$ & \centering $3.36$ & \centering $57.9$ & \centering $0.24$ & \ \ $0.96$ \\
\hline \centering $30$ & \centering $0.104$ & \centering $2.69$ & \centering $25.85$ & \centering $0.33$ & \ \ $0.526$ \\
\hline \centering $45$ & \centering $0.14$ & \centering $2.43$ & \centering $17.62$ & \centering $0.4$ & \ \ $0.321$ \\
\hline \centering $70$ & \centering $0.15$ & \centering $2.43$ & \centering $14.17$ & \centering $0.43$ & \ \ $0.194$ \\
\hline \centering $100$ & \centering $0.17$ & \centering $2.43$ & \centering $16.08$ & \centering $0.45$ & \ \ $0.134$ \\
\hline
\end{tabular}
\end{center}
\caption[]{Critical parameters which separate stable and unstable
regions  (tangent $k_0$, wave number $\alpha_0$ and
$1/\varepsilon_0$); and $Q$ and $R$ are functions of
  $\Delta\Phi$.} \label{T1}
\end{table}
The initial point $(1/\varepsilon, \alpha)$ belonging to region I
corresponds to decaying perturbations, but with
increasing time the instantaneous values of $(1/\varepsilon, \alpha)$
intersect the lower branch of the neutral curve in point $1$ and
enter the instability region. For $(1/\varepsilon, \alpha)$ in
region II, the perturbations are increasing until the
instantaneous values $(1/\varepsilon, \alpha)$ intersect the upper
branch of the neutral curve; after this instant moment they begin
to decay.
For $(1/\varepsilon, \alpha)$ in region III, the
perturbations decrease  monotonically at all times.

It is possible to show that for $k<k_0$ the straight lines always
leave the instability region. Indeed, from \eqref{eqDelta F}, \eqref{eq87P} in combination with \eqref{eq88e}, \eqref{eq88c}
the asymptotics for the V--C curve as $\varepsilon \to 0$ can be obtained
$$
\eta_m^3\sim\frac{9\sqrt{\pi}}{32}(\varepsilon \Delta \Phi)^2
$$
just from this relation
\begin{equation}\label{eq15212}
r=\alpha \eta_m \sim \left(\frac{9\sqrt{\pi}\Delta \Phi^2}{32}\right)^{1/3}\alpha \varepsilon^{2/3}.
\end{equation}
Now the upper
branch of the marginal stability curves in short--wave approximation,
$\alpha \to \infty$, can be readily obtained. In particular,
we can claim that
\begin{equation}\label{Q}
\alpha \sim Q(\Delta \Phi)\left(\frac{1}{\varepsilon}\right)^{2/3},
\end{equation}
where the function $Q(\Delta \Phi)$ is calculated numerically and
presented in  Table~\ref{T1}.

In order to evaluate the lower marginal stability branch, it is  supposed that
\begin{equation}\label{R}
\alpha\sim R(\Delta\Phi).
\end{equation}
At $\varepsilon\to0$ the relation becomes exact. The dependence $R(\Delta\Phi)$ is tabulated in  Table~\ref{T1}.

The relations \eqref{Q} and \eqref{R} allow, in particular, to
estimate the time gap $\Delta \tilde{t}$ in which the perturbation
with $k<k_0$ belongs  to the instability region, between points $1$
and $1'$, in Fig.~\ref{Alpeps},
$$
\Delta \tilde{t} = \tilde{t}'-\tilde{t} \sim \frac{k^4Q^6-R^2}{4\tilde{\alpha}^2\tilde{D}}.
$$
This relation becomes exact at $k=\tilde{\alpha}\tilde{\lambda}_D
\to 0$.

\section{CONCLUSIONS AND DISCUSSION}

Electro--convective processes in an electrolyte solution between a
semi--selective ion--exchange  membranes have been investigated
numerically and asymptotically in the limit of small Debye length $\varepsilon$.
First,  a simplified system of equations has been
derived from the full system of equations that describes the ion transport and contains   the
Poisson and  Stokes equations. Asymptotic expansions in
small dimensionless Debye length are applied to both equilibrium
and stability problems  for the limiting regimes. Only the
outer solution has been considered, ignoring the input of the surface
and internal boundary layers. A novel class of the 1D unsteady
self--similar solutions is found. A linear stability of these
solutions and transition to electro--convection is considered. The
marginal stability curves are constructed both numerically and asymptotically.

 The results of stability modelling in small $\varepsilon$
 are in fair agreement with those
obtained by direct numerical simulations of the entire problem
everywhere except for the vicinity of the critical point of the
marginal stability curves. The possible reason of that deviation
is  that the critical point of the marginal curves is
located near the transition region from the under--limiting to limiting regimes, while our asymptotics is valid only for the limiting
regimes.

{\bf{ACKNOWLEDGMENTS}}

E.A.D. is grateful to  I.Rubinstein for his stimulating questions
and remarks and to  V. V. Nikonenko for numerous discussions and
significant help in understanding  the physical pattern  the
process.

\newpage

\renewcommand{\refname}{References}

\newpage

{\bf Figure Captions}
\begin{itemize}

\item[\bf Figure 1] Schematic V--C curve for ion--exchange  membranes. A, B and C
represent  under--limiting,  limiting and over--limiting regimes,
respectively;
* depicts the threshold of the
over--limiting regimes; $C_I$ and $C_{II}$ are the regions of
regular and irregular current oscillations; $j$ and $j_{lim}$ are
the current density at the bottom membrane and
its limiting value; $\Delta V$ is the potential drop.

\item[\bf Figure 2]
Schematic for limiting processes in a steady--state
equilibrium: a)  electric field $E{\,=\,}d \Phi / dy$; b)
volume charge density $\rho{\,=\,c^+\,-\,}c^-$; c) ion
concentrations $c{\,=\,}c^+$ and $c{\,=\,}c^-$. The dashed line is
the outer solution  in the small vicinities of $y{\,=\,}0$ (the
double ion boundary layer) and $y{\,=\,}y_m$ (the boundary layer
between the volume space charge region and the diffusion layer).
The solid and dashed lines depict  exact and  asymptotic solutions
for small $\varepsilon$.

\item[\bf Figure 3] Marginal stability curves at different $\varepsilon$,
numerics. The under--limiting regimes are  stable (adopted from
Demekhin et al. \cite{Dem}).

\item[\bf Figure 4] Self--similar charge $\rho=c^+-c^-$ (a) and concentration
$c^+$ distribution (b) for different $\varepsilon$. Curves correspond to the following values of electric potential drop: 1. $\Delta \Phi=50$; 2. $\Delta \Phi=100$; 3. $\Delta \Phi=150$; 4. $\Delta \Phi=200$.

\item[\bf Figure 5] VC curves. Comparison of the self--similar asymptotics (dashed lines) and the exact numerics (solid lines) for several $\varepsilon$.

\item[\bf Figure 6] Distribution of the charge density $\rho$ in space for several moments of time, $\Delta V=100$ and $\nu{\,=\,}\tilde{\lambda}_D/\tilde{L}=0.001$. Solid line corresponds to the numerical solution of the unsteady problem \eqref{eqN1111}--\eqref{eq2770P111}. Two--parametric self--similar solution of \eqref{eq41}--\eqref{eq44} with $\varepsilon$ and $\Delta \Phi$ parametrically changing with time is depicted by dashed lines.

\item[\bf Figure 7] Typical evolution of $1/j(t)$. I is the short--time region of influence of the initial data; II is the intermediately--long--time
region of self--similarity, and III is the region of influence of the upper membrane $y{\,=\,}1$, ($\Delta V{\,=\,}100$ and $\nu{\,=\,}\tilde{\lambda}_D/\tilde{L}{\,=\,}0.001$).

\item[\bf Figure 8] VC curves. Solid line relates to \eqref{eq88dd}, while  dashed line -- to the simplified version \eqref{eq88c}.

\item[\bf Figure 9] Negative ion concentration (a), $c^-$, and electric field (b), $H$, vs $y$ for several moments of time. Comparison of numerics
\eqref{eqN111}--\eqref{eqJPDim11} (solid lines) with self--similar asymptotics (\ref{eq82})--(\ref{eq85b}) (dashed lines) for (a) $\Delta V{\,=\,}50$, $\nu{\,=\,}0.001$: 1. $t{\,=\,}0.0001$; 2. $t{\,=\,}0.001$; 3. $t{\,=\,}0.004$; 4. $t{\,=\,}0.015$; 5. $t{\,=\,}0.5$;  and  (b) $\Delta V{\,=\,}100$, $\nu{\,=\,}0.0005$: 1. $t{\,=\,}0.001$; 2. $t{\,=\,}0.003$; 3. $t{\,=\,}0.03$.

\item[\bf Figure 10] Comparison of the V--C characteristics for
self--similar solution and numerics, $\Delta V=50, \nu = 0.00005$.
Numerics \eqref{eqN111}--\eqref{eqJPDim11} is shown by squares and
the universal self--similar VC curve \eqref{eq88dd} -- by solid
line. The region of the solution self--similarity  $t_1{\,<\,t\,<\,}t_2$,
$t_1\gg\nu^2/4 $, $t_2\ll t_s=\tilde{L}^2/(4 \tilde{D})$.
Arrows depict the time growth along the voltage--current curve.

\item[\bf Figure 11] Numerical points for several values of the potential drop
between the membranes, $\Delta V$, shrink into the universal  VC curve $J(\Delta F)$ given by
\eqref{eq88dd}; a) $\nu{\,=\,}0.001$, b) $\nu{\,=\,}0.0005$.

\item[\bf Figure 12] Components of scaled slip velocities $u$ and $v$.

\item[\bf Figure 13] Stream--lines for $\Delta\Phi{\,=\,}100,\ \varepsilon{\,=\,}0.001$ and different values of $\alpha$.

\item[\bf Figure 14] Marginal stability curves for self--similar asymptotic in $\varepsilon$ solution \eqref{eq1515}--\eqref{eq1516} (dashed lines) and exact numerical solution (solid lines) \cite{Dem}.

\item[\bf Figure 15] Marginal stability curves in $1/\varepsilon$ and $\alpha$ coordinates.

\item[\bf Figure 16] The universal functions a) $G{\,=\,}G(r)$, b) $a{\,=\,}a(r)$.

\end{itemize}

\newpage

{\bf Table Caption}
\begin{itemize}

\item[\bf Table] Critical parameters which separate stable and unstable
regions  (tangent $k_0$, wave number $\alpha_0$ and
$1/\varepsilon_0$); and $Q$ and $R$ are functions of
$\Delta\Phi$.

\end{itemize}
\newpage
\appendix
\subsection{ Electrostatic solution  in the space charge region}

Substituting a new independent variable $\displaystyle\xi=\alpha\left(\eta_m-\eta\right)$ into
\eqref{eq1123}--\eqref{eq1124} results in:
\begin{equation}\label{eq1312}
\displaystyle
\xi^\frac{1}{2}\frac{d^2}{d\xi^2}\left(\xi^\frac{1}{2}\frac{d \hat{F}}{d\xi}\right)-\xi\frac{d\hat{F}}{d\xi}-\hat{F}=0
\end{equation}
\begin{equation}\label{eq1313}
\displaystyle  \ \ \hat{F}=0\ \ \mbox{for} \ \ \ \xi=0
\end{equation}
\begin{equation}\label{eq1314}
\displaystyle \ \ \hat{F}=0 \ \ \mbox{for} \ \ \ \xi=r.
\end{equation}
where $r = \alpha\eta_m$. The additional conditions which
originate from \eqref{eq1124} are:
\begin{equation}\label{eq1315}
\displaystyle\sqrt{\frac{2\alpha}{J}}\xi^\frac{1}{2}\frac{d\hat{F}}{d\xi}\to-\hat\eta_m\ \ \ \mbox{at}\ \ \ \xi\to 0,
\end{equation}
\begin{equation}\label{eq1316}
\displaystyle\alpha\sqrt{\frac{J}{2\eta_m}}\left(2r\frac{d^2\hat{F}}{d\xi^2}+\frac{d\hat{F}}{d\xi}\right)=-\hat{J}\ \ \ \mbox{for}\ \ \ \xi=r.
\end{equation}
We present the solution of \eqref{eq1312}--\eqref{eq1314} as
\begin{equation}\label{eq1317}
\displaystyle
\hat{F}=B\sqrt{\xi}f(\xi)
\end{equation}
where $B$ is constant.  The equation \eqref{eq1312} turns
into
$$\xi f'''+\frac{5}{2} f''-\xi f'-\frac{3}{2} f=0.$$
 The general regular solution of this equation, up to a
constant factor, has the form $\hat{F}=B\sqrt{\xi}\left(
f_1\left(\xi\right)+a f_2\left(\xi\right) \right)$, where $f_1$
and $f_2$ are two linearly independent solutions of the equation,
$a$ is determined below. Solutions  $f_1$, $f_2$,  can be
easily found numerically. Instead, for comparison convenience with
long--wave asymptotics, the solution is presented as power
expansions
\begin{equation}\label{eq1328}
\displaystyle
f_1\left(\xi\right)=\sum_{n=0}^\infty{a_{2n}\xi^{2n}},\ \ \
\displaystyle
f_2\left(\xi\right)=\sum_{n=0}^\infty{a_{2n+1}\xi^{2n+1}},
\end{equation}
where  $\displaystyle a_k \sim1/k!$, and the series  have infinite
radius of convergence
\begin{equation}\label{eq1330}
\begin{array}{c}
a_{2n}=\displaystyle \frac{1}{(2n)!}\prod_{k=1}^{n}\frac{4k-1}{4k+1},\ \ \ a_0=1,\\[12pt]
a_{2n+1}=\displaystyle \frac{1}{(2n+1)!}\prod_{k=1}^{n}\frac{4k+1}{4k+3},\ \ \ \ a_1=1.
\end{array}
\end{equation}
Note that the long wave limit is described by a finite term
expansion \eqref{eq1328}.

Assuming
$$
\displaystyle f = f_1+a f_2=0
$$
at some $\alpha\eta_m =  r\left(a\right)$, the
second BC \eqref{eq1314} for $\hat{F}$ is satisfied. Let us invert the
function $r=r(a)$,
\begin{equation}\label{a}
a(r)=-\frac{f_1(r)}{f_2(r)},\ \ \ r=\alpha\eta_m.
\end{equation}

Thus we obtain, taking into account \eqref{eq1317}, that the
function  $\hat{F}$ satisfies the first two boundary
conditions \eqref{eq1313}--\eqref{eq1314}. The constant  $B$ is
obtained by substitution of \eqref{eq1317} into \eqref{eq1316},
\begin{equation}\label{eq1321}
\hat\eta_m=-\frac{G\left(r\right)}{\alpha}\frac{\hat
J}{J},\ \ \ G(r)=-\frac{1}{2rf''(r)+3f'(r)}.
\end{equation}
where $f',\ f''$ are taken at $\xi=r$ and are functions of $r$, $G
= G\left(r\right)$ is the universal function shown along with
$a(r)$ in Fig.~\ref{Ga}.

\begin{figure}
  \begin{center}
    \includegraphics[height=6cm]{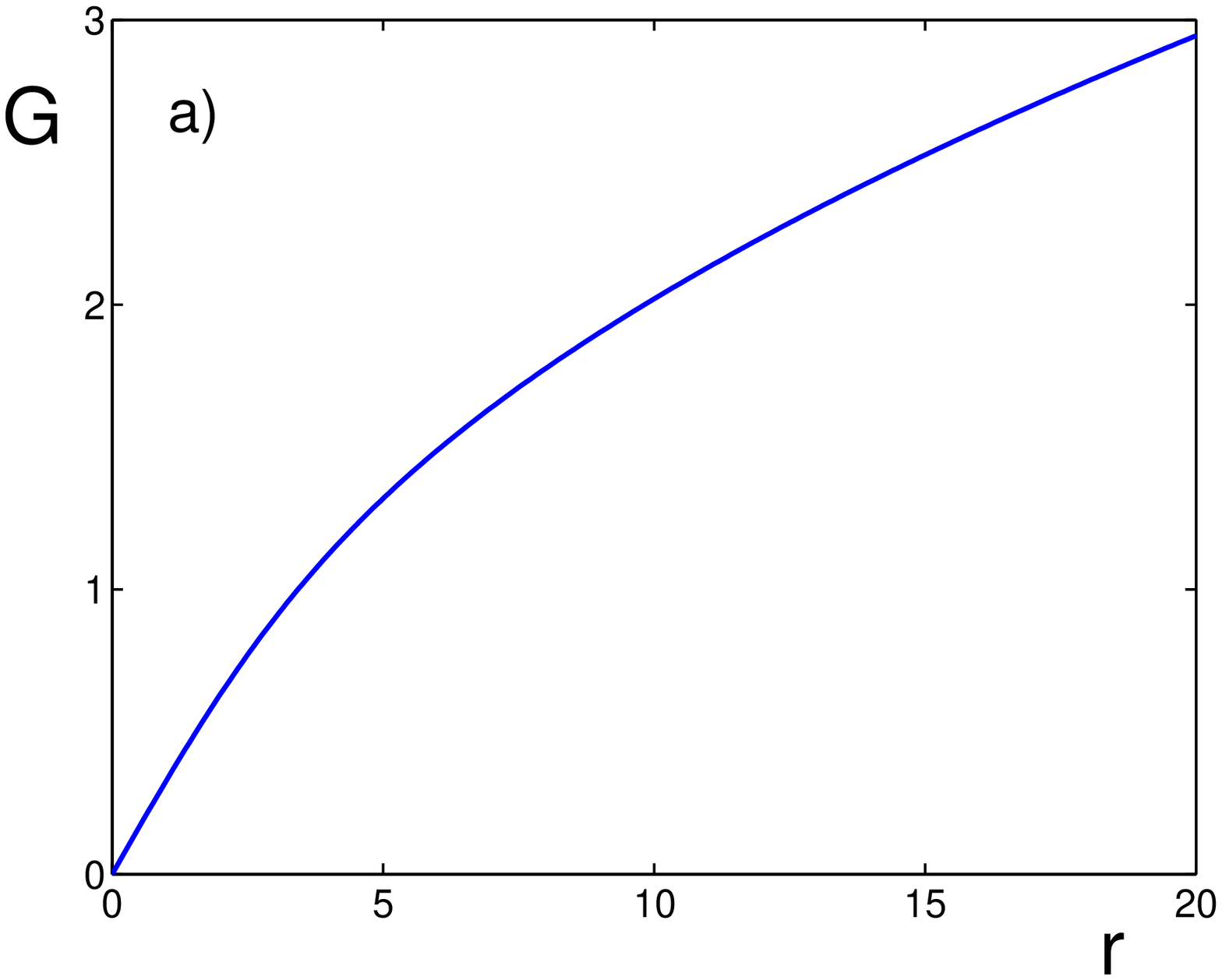}
    \includegraphics[height=6cm]{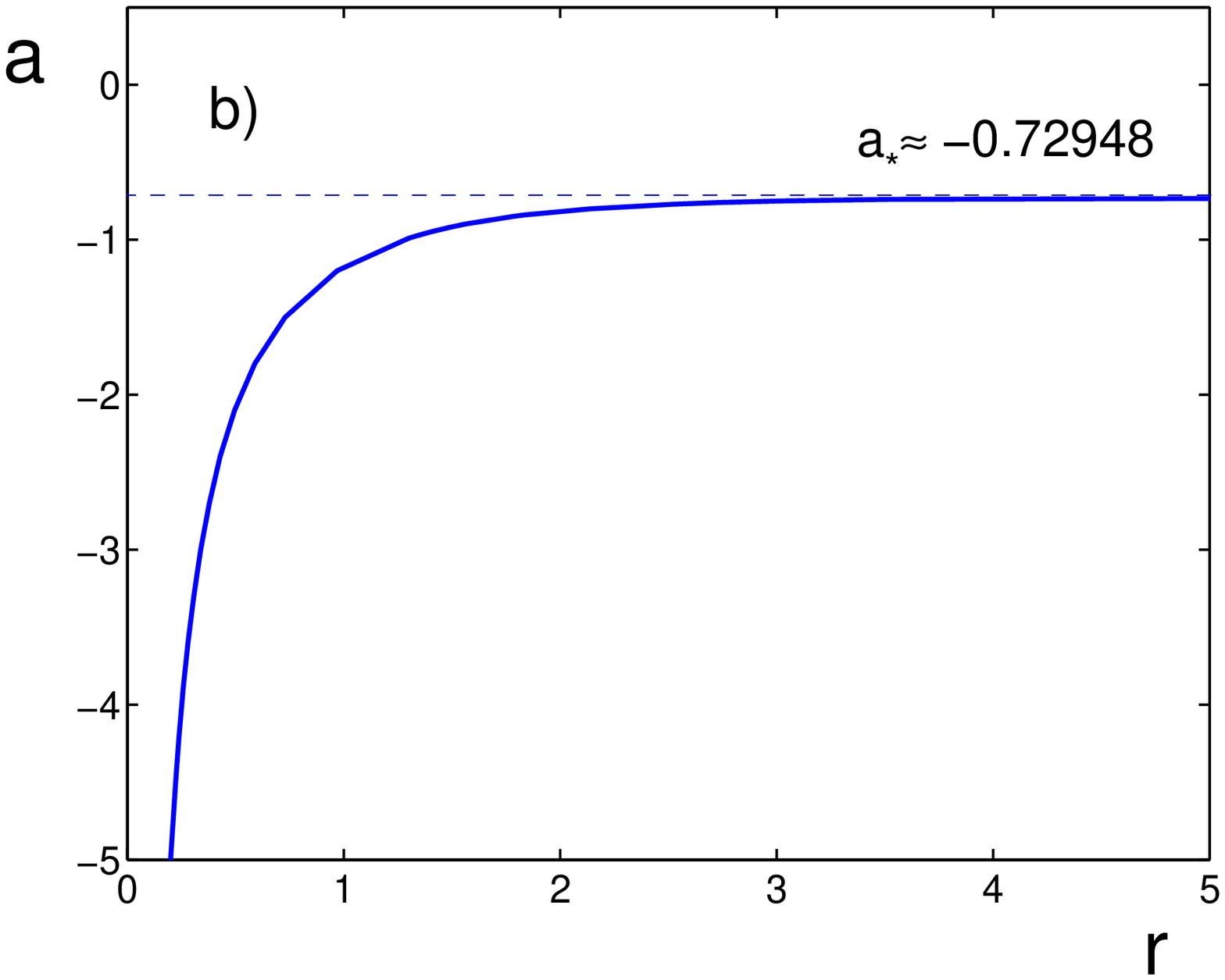}
  \end{center}
\caption{The universal functions a) $G{\,=\,}G(r)$, b) $a{\,=\,}a(r)$.}\label{Ga}
\end{figure}

Then we get the unknown function $\hat{F}$,
\begin{equation}\label{eq1322}
\hat{F}=-\sqrt{\frac{2J}{\alpha}}\:\hat{\eta}_m\sqrt{\xi}f(\xi),\ \ \ \xi=\alpha\left(\eta_m-\eta\right).
\end{equation}
Substituting  \eqref{eq1322} into \eqref{distRho}, we obtain the
relation for the outer expansion of the charge density $\hat{\rho}$,
\begin{equation}\label{eq132201}
\hat{\rho}=\varepsilon\sqrt{2J\alpha^3}\:\hat{\eta}_m\left[\sqrt{\xi}f\:''+\frac{1}{\xi}f\:'-\left(\frac{1}{4\xi\sqrt{\xi}}+\sqrt{\xi}\right)f\right].
\end{equation}
which  has a singularity at $\eta=\eta_m$.

Furthermore, from  \eqref{eq1328}--\eqref{a}, the asymptotic
expansions for $a(r)$ and $G(r)$ as $r\rightarrow0$ can be found
\begin{equation}\label{as}
a(r) =
-\frac{\sum_{n=0}^\infty{a_{2n}r^{2n}}}{\sum_{n=0}^\infty{a_{2n+1}r^{2n+1}}}.
\end{equation}
For small $r$, which corresponds to the long--wave approximation
($\alpha\rightarrow0$),we  obtain,
\begin{equation}\label{as10}
a\left(r\right) = -\frac{1}{r} -
\frac{19}{105}r+\frac{169}{24255}r^3+\cdots \sim-\frac{1}{r},\ \
r\rightarrow0.
\end{equation}
From \eqref{eq1321}, in the first approximation with respect to
$r$,
\begin{equation}\label{as11}
G\left(r\right)=\frac{1}{3}\,r-\frac{1}{210}\,r^3+\frac{59}{415800}\,r^5+\cdots\sim\frac{1}{3}\,r,\
\ r\rightarrow0.
\end{equation}
and
\begin{equation}\label{eq1334a}
\hat\eta_m=\eta_m\frac{{\hat J}}{J}\left(-\frac{1}{3}+\frac{1}{210}(\alpha\eta_m)^2-\frac{59}{415800}(\alpha\eta_m)^4+\cdots\right)\sim-\frac{\eta_m}{3}\frac{{\hat J}}{J}.
\end{equation}
Approximation \eqref{as11}--\eqref{eq1334a} is valid at
$r<1.5\div2$.

The asymptotics of $f_1$ and $f_2$ at $\xi\rightarrow\infty$ are
useful in the short--wave approximation $\alpha \to \infty$  of the
stability problem. They can be readily found by asymptotic
expansions of the equation \eqref{eq1312}:
$$
\displaystyle f_1 \sim \frac{const_1}{\sqrt{\xi}}\:e^\xi, \ \ \ \ f_2 \sim
\frac{const_2}{\sqrt{\xi}}\:e^\xi,\ \ \ \ \xi\rightarrow\infty.
$$
At  $r\rightarrow\infty$
\begin{equation}\label{G}
a(r) \sim a_* ,\ \ \
G\left(r\right)\sim c_*\sqrt{r},
\end{equation}
where
$$
a_*\approx-0.72948,\qquad c_*\approx0.66468.
$$

\subsection{Components of the slip velocity}

Now the obtained function $\hat F$ can be used to solve the
hydrodynamic  problem \eqref{eq1127}, \eqref{BC_0}, \eqref{BC_eta_m} and find
the stream function perturbation $\hat{\Psi}$. The slip velocity
components $\hat\Psi\left(\eta_m\right)=i\hat{V}_m$ and
$\hat\Psi\:\left(\eta_m\right)=\hat{U}_m$ specify the boundary
conditions for the problem in the electro--neutral region $\eta_m <
\eta <\infty$.

Taking the variable $\xi=\alpha(\eta_m-\eta)$
with $r=\alpha\eta_m$, and utilizing
relations \eqref{eq1321}, \eqref{eq1322}, the problem
\eqref{eq1127}--\eqref{BC_eta_m}  can be reformulated as
\begin{equation}\label{eq1400a}
\displaystyle
\frac{d^4\psi}{d\xi^4}-2\frac{d^2\psi}{d\xi^2}+\psi=\xi\left(\frac{d^2
f}{d \xi^2}-f\right)+\frac{d f}{d \xi},
\end{equation}
\begin{equation}\label{eq1401a}
\begin{array}{c}
\displaystyle\frac{d^2\psi}{d\xi^2}-2\frac{d\psi}{d\xi}+\psi = 0,\ \ \ \frac{d^3\psi}{d\xi^3}-3\frac{d\psi}{d\xi}+2\psi = 0\ \ \ \mbox{for}\ \ \ \xi=0,\\
\displaystyle\psi=\frac{d\psi}{d\xi}=0\ \ \ \mbox{for}\ \ \ \xi=r,
\end{array}
\end{equation}
where
\begin{equation}\label{psia}
\psi\left(\xi,r\right)=\frac{1}{i\hat C}\hat{\Psi}, \ \ \ C=\frac{2\varkappa}{\varepsilon^2\alpha^4}\:G\left(r\right)\hat{J}.
\end{equation}
$\hat{V}_m$ and $\hat{U}_m$ are used as the boundary condition for the hydrodynamic problem  in the
electro--neutral region  $\eta > \eta_m$,
\begin{equation}\label{as3a}
\hat{V}_m=\hat Cv\left(r\right) \equiv \hat C\psi\left(0,r\right),\ \ \
\hat{U}_m=-i\alpha \hat Cu\left(r\right) \equiv -i\alpha \hat C
\frac{\partial\psi}{\partial\xi}\left(0,r\right).
\end{equation}

The function $f$ in the right hand side of \eqref{eq1400a} is
$$
\displaystyle \xi\left(\frac{d^2 f}{d \xi^2}- f\right)+ \frac{d
f}{d \xi} \stackrel{def}{=}g=g_1+ag_2,
$$
where
\begin{equation}\label{psio}
\begin{array}{c}
\displaystyle g_1=\sum_{n=0}^{\infty}b_{2n+1}\xi^{2n+1},\ \ \ \ b_{2n+1}=\frac{a_{2n}}{\left(2n+1\right)\left(4n+5\right)},\\
\displaystyle
\displaystyle g_2=\sum_{n=0}^{\infty}b_{2n}\xi^{2n},\ \ \ \ b_{2n}=\frac{a_{2n-1}}{2n\left(4n+3\right)},\ \ b_0=1.
\end{array}
\end{equation}
Then the solution of \eqref{eq1400a}\textbf{--}\eqref{eq1401a} can be presented in the form
\begin{equation}\label{psiseries}
\psi=(A+B\xi)e^\xi+\psi_p\:,\ \ \ \psi_p=\psi_1+a\psi_2\:,
\end{equation}
where $e^\xi$ and $\xi e^\xi$ \textbf{--} are two linearly independent
solutions which satisfy the first two conditions \eqref{eq1401a} at $\xi{\,=\,}0$, $A$, $B$ \textbf{--} some constants, $\psi_p$ is a particular solution of \eqref{eq1400a} which satisfies zero conditions
\begin{equation}\label{unif}
\psi_p = \frac{d\psi_p}{d\xi} = \frac{d^2\psi_p}{d\xi^2} = \frac{d^3\psi_p}{d\xi^3} = 0,\ \ \ \xi=0.
\end{equation}
Functions $\psi_1$ and $\psi_2$ \textbf{--} are the odd and even components of $\psi_p$, which can be found from solutiion \eqref{eq1400a} with right--hand side $g_1$ and $g_2$ correspondingly and boundary conditions~\eqref{unif}. They can be presented as expansions
\begin{equation}\label{psioe}
\begin{array}{ll}
\displaystyle\psi_1=\sum_{n=2}^{\infty}c_{2n+1}\xi^{2n+1}, &\displaystyle\ \ \  c_{2n+1}=\sum_{k=0}^{n-2}\frac{\left(n-k-1\right)\left(2k+1\right)!}{\left(2n+1\right)!}\,b_{2k+1},\\[12pt]
\displaystyle
\psi_2=\sum_{n=2}^{\infty}c_{2n}\xi^{2n}, & \displaystyle\ \ \ c_{2n}=\sum_{k=0}^{n-2}\frac{\left(n-k-1\right)\left(2k\right)!}{\left(2n\right)!}\,b_{2k}
\end{array}
\end{equation}
The constants  $A, B$ are found from the BC's at $\xi=r$,
\begin{equation}\label{AB}
\displaystyle A=\left(r \frac{d \psi_p(r)}{d
\xi}-(1+r)\psi_p(r)\right)e^{-r},\ \ \ \ B=\left(\psi_p(r)-\frac{d
\psi_p(r)}{d \xi}\right)e^{-r}.
\end{equation}
Taking into account \eqref{as3a} yields
\begin{equation}\label{u}
v = A = \left(r\frac{d \psi_p(r)}{d
\xi}-(1+r)\psi_p(r)\right)e^{-r},
\end{equation}
\begin{equation}\label{v}
u = A+B = \left((r-1)\frac{d \psi_p(r)}{d
\xi}-r\psi_p(r)\right)e^{-r},
\end{equation}
The obtained relations provide a complete description for the slip
velocity components $\hat{U}_m$ and $\hat{V}_m$  according to
\eqref{as3a}.

Finally, let us present the Taylor series near $r = 0$ for the slip
velocity in the long--wave approximation.
Using \eqref{as10}, \eqref{psiseries} and \eqref{psioe},
we can write  the following relation,
\begin{equation}\label{eq1418}
\psi_p\left(r\right)= \sum_{k=1}^{\infty}c_{2k+3} r^{2k+3}- \frac{\sum_{k=0}^{\infty}a_{2k}r^{2k}}{\sum_{k=0}^{\infty}a_{2k+1}r^{2k+1}}\sum_{k=1}^{\infty}c_{2k+2}r^{2k+2},
\end{equation}
where the series coefficients are calculated according to
\eqref{eq1330} and \eqref{psio}, \eqref{psiseries}. Expanding the
exponents in \eqref{u}, \eqref{v} in series in the vicinity of
$r=0$, re--expanding them and differentiating the above relation
with respect to $r$,  the functions  $u(r)$ and $v(r)$ are obtained,
\begin{equation}\label{vs}
\begin{array}{l}
\displaystyle v(r) =
-\frac{1}{8}r^3+\frac{1}{6}r^4-\frac{27}{200}r^5 + \cdots,\\[7pt]
\displaystyle u(r) = \frac{1}{6}r^2-\frac{1}{3}r^3+\frac{167}{504}r^4 + \cdots.
\end{array}
\end{equation}
All the series in \eqref{eq1418} converge
at any $r$, while the series \eqref{vs} have finite convergence
radius because of above mentioned hidden singularity of the universal function $a=a(r)$ in the complex plane.

We find the limiting values of  $u$ and $v$,
\begin{equation}\label{uvinfty}
v\rightarrow v_{\infty}\approx-0.27245,\ \ \ \ u\rightarrow
u_{\infty}\approx-0.12437\ \ \ \ \mbox{as}\ \ r\rightarrow\infty.
\end{equation}
Moreover, $u$ changes sign  at $r=\alpha\eta_m\approx1.59523$. The
dependence of $u$ and  $v$ on $r$ is shown in Fig.~\ref{uv}. At $r
\to 0$ results can be obtained in a closed form; they coincide
with obtained in \cite{Rub3a}.

\end{document}